%% file: main.tex
\documentclass[preprint,5p]{elsarticle}
\input{preamble}

\begin{document}
    
\begin{frontmatter}
\title{Generic and Robust Root Cause Localization for Multi-Dimensional Data in Online Service Systems}

\author[1,6]{Zeyan Li}
\ead{zy-li18@mails.tsinghua.edu.cn}

\author[2]{Junjie Chen}
\ead{junjiechen@tju.edu.cn}

\author[1]{Yihao Chen}
\ead{chenyiha17@mails.tsinghua.edu.cn}

\author[1]{Chengyang Luo}
\ead{luocy16@mails.tsinghua.edu.cn}

\author[1]{Yiwei Zhao}
\ead{zhaoyw17@mails.tsinghua.edu.cn}

\author[3]{Yongqian Sun}
\ead{sunyongqian@nankai.edu.cn}

\author[4]{Kaixin Sui}
\ead{suikaixin@bizseer.com}

\author[1]{Xiping Wang}
\ead{wxp17@mails.tsinghua.edu.cn}

\author[4]{Dapeng Liu}
\ead{liudapeng@bizseer.com}

\author[5]{Xing Jin}
\ead{jinxing.zh@ccb.com}

\author[5]{Qi Wang}
\ead{wangqi10.zh@ccb.com}

\author[1,6]{Dan Pei}
\ead{peidan@tsinghua.edu.cn}

\affiliation[1]{
    organization={Tsinghua University},
    city={Beijing}, 
    country={China}
}

\affiliation[2]{
    organization={Tianjin University, College of Intelligence and Computing},
    city={Tianjin}, 
    country={China}
}

\affiliation[3]{
    organization={Nankai University, College of Software},
    city={Tianjin}, 
    country={China}
}

\affiliation[4]{
    organization={Bizseer},
    city={Beijing}, 
    country={China}
}

\affiliation[5]{
    organization={China Construction Bank},
    city={Beijing}, 
    country={China}
}

\affiliation[6]{
    organization={Beijing National Research Center for Information Science and Technology (BNRist)},
    city={Beijing}, 
    country={China}
}

\begin{abstract}
Localizing root causes for multi-dimensional data is critical to ensure online service systems' reliability.
When a fault occurs, only the measure values within speciﬁc attribute combinations (e.g., Province=Beijing) are abnormal. 
Such attribute combinations are substantial clues to the underlying root causes and thus are called root causes of multi-dimensional data.
This paper proposes a generic and robust root cause localization approach for multi-dimensional data, PSqueeze.
We propose a generic property of root cause for multi-dimensional data, generalized ripple effect (GRE).
Based on it, we propose a novel probabilistic cluster method and a robust heuristic search method.
Moreover, we identify the importance of determining external root causes and propose an effective method for the first time in literature. 
Our experiments on two real-world datasets with 5400 faults show that the F1-score of PSqueeze outperforms baselines by 32.89\%, while the localization time is around 10 seconds across all cases.
The F1-score in determining external root causes of PSqueeze achieves 0.90. 
Furthermore, case studies in several production systems demonstrate that PSqueeze is helpful to fault diagnosis in the real world.
\end{abstract}

\begin{keyword}
Root cause localization \sep online service system \sep ripple effect \sep multi-dimensional data

\end{keyword}

\end{frontmatter}

\input{introduction}
\input{background}
\input{gre}
\input{methodology}
\input{evaluation}
\input{deployment}

\input{discussion}

\input{related_work}

\input{conclusion}

\section{Acknowledgment}
The authors gratefully thank  Suning.com Co., Ltd. for providing parts of the data used in this paper's evaluation. 
We thank Juexing Liao and Chuanxi Zheng for proofreading this paper. 
This work is supported by the National Key R\&D Program of China 2019YFB1802504, and the State Key Program of National Natural Science of China under Grant 62072264.

\bibliographystyle{elsarticle-num}
\bibliography{references}
\end{document}

%% file: preamble.tex
\usepackage{amsthm}
\usepackage{amsmath}
\usepackage{amssymb}
\usepackage{array}
\usepackage{color,colortbl}
\usepackage{multirow}
\usepackage{xcolor}
\usepackage{float}
\usepackage{tabularx}
\usepackage{url}
\usepackage{algorithm}
\usepackage[noend]{algpseudocode}
\usepackage[OT1]{fontenc}
\usepackage{placeins}
\usepackage{wrapfig}
\usepackage{listings}
\usepackage[caption=false,font=footnotesize,labelfont=sf,textfont=sf]{subfig}
\usepackage{cleveref}
\usepackage{enumitem}
\usepackage{sistyle}
\usepackage{makecell}
\usepackage{adjustbox}
\usepackage[utf8]{inputenc}%
\usepackage{longtable}
\usepackage{booktabs}
\usepackage{tcolorbox}

\allowdisplaybreaks[4]

\SIthousandsep{,}

\newcolumntype{Y}{>{\centering\arraybackslash}X}

\newcommand{\ours}[0]{\textit{PSqueeze}}

\newcommand{\eg}[0]{\textit{e.g.}}
\newcommand{\ie}[0]{\textit{i.e.}}
\newcommand{\vv}[1]{\boldsymbol{#1}}
\newcommand{\RNum}[1]{#1}

\newcommand{\ds}[1]{$\mathcal{#1}$}

\newcommand{\greycell}[0]{\cellcolor[HTML]{DEDEDE}}

\AtBeginDocument{
\crefformat{equation}{#2(#1#3)}
\crefformat{section}{#2Section~#1#3}
\crefformat{subsection}{#2Section~#1#3}
\crefformat{subsubsection}{#2Section~#1#3}
\crefformat{algorithm}{#2Algorithm~#1#3}
\crefformat{figure}{#2Fig.~#1#3}
\crefformat{subfigure}{#2Fig.~#1#3}
\crefformat{table}{#2Table~#1#3}
}

\newcounter{conclusion}[section]    
\newenvironment{conclusion}
{
\begin{tcolorbox}[boxsep=0pt, left=1pt, right=1pt, top=2pt, bottom=2pt]
\textbf{Finding~\refstepcounter{conclusion}\theconclusion}
}    
{ 
\end{tcolorbox}
}

%% file: introduction.tex
\section{Introduction}
\label{sec:introduction}

Large online service systems (\eg{}, online shopping platforms) serve millions of users and require high reliability to ensure user experience.
Faults in large online service systems could cause enormous economic loss and damage user satisfaction~\cite{chen2019empirical}.
For example, the loss of one-hour downtime for \texttt{Amazon.com} on Prime Day in 2018 (its biggest sale event of the year) is up to \$100 million~\cite{2018amazon}.
Therefore, it is in urgent demand to diagnose faults rapidly.

To ensure quality of software service, operators usually closely monitor some measures (\eg{}, total dollar amount), which reflect the system status~\cite{lin2020fast,sun2018hotspot}.
When a fault occurs, the monitoring system can detect the abnormal measure values and raise alerts to operators.
A measure record is associated with many attributes, and when a fault occurs, only the measure values of specific attribute combinations are abnormal~\cite{lin2020fast,sun2018hotspot,gu2020efficient}.
For example, when the network service provided by \textit{CMobile} in \textit{Beijing} province fails, only the dollar amount of (Province=Beijing, ISP=CMobile) would decrease dramatically.
Such \textit{attribute combinations} can effectively indicate the fault location and serve as substantial clues to the underlying root causes~\cite{lin2020fast,gu2020efficient}.
Thus, we call the set of such attribute combinations as the \textit{root cause} of the multi-dimensional data~\cite{sun2018hotspot,li2019generic,lin2020fast,ahmed2017detecting}.
Therefore, following existing work~\cite{lin2016idice,lin2020fast,gu2020efficient,rong2020locating,sun2018hotspot,li2019generic,ahmed2017detecting}, in this paper, we focus on \textit{localizing root causes of multi-dimensional data} to help operators diagnose faults rapidly.

However, it is challenging due to the huge search space.
On the one hand, there are many attributes (\eg{}, dozens) and attribute values (\eg{}, thousands) in large online service systems, leading to a combinatorial explosion.
On the other hand, faults must be mitigated rapidly to reduce the impact on user experience, and thus it requires high efﬁciency for the localization.

Existing works~\cite{bhagwan2014adtributor, persson2018anomaly,lin2016idice,sun2018hotspot,ahmed2017detecting,gu2020efficient,lin2020fast,rong2020locating} apply various techniques to overcome the huge search space challenge, but they are not generic or robust enough due to some limitations or not efficient enough (see later in \cref{tbl:related-works}).
For example, MID~\cite{gu2020efficient}, iDice~\cite{lin2016idice}, and ImpAPTr~\cite{rong2020locating} are only applicable to specific types of measures.
Apriori~\cite{ahmed2017detecting,lin2020fast} and R-Adtributor~\cite{persson2018anomaly} highly relies on parameter fine-tuning.
Notably, all the previous approaches do not check \textit{external root causes}, \ie{}, root causes containing some unrecorded or unused attributes.
The localization results are always incorrect when there are external root causes, which can mislead the direction of fault diagnosis and waste time~\cite{kim2013root}.

This paper proposes \ours{}, a generic and robust root cause localization approach for multi-dimensional data.
Rather than impractical root cause assumptions or properties of specific kinds of measures, the search strategies of \ours{} are based on a more generic property of root causes of multi-dimensional data, \textit{generalized ripple effect} (GRE).
GRE holds for different measures (see \cref{sec:generalized-ripple-effect}) and holds in real-world faults (see \cref{sec:deployment}), enabling our search strategies' genericness.
Based on GRE, we propose a ``bottom-up\&top-down'' method to achieve high efficiency without much loss of genericness and robustness.
Specifically speaking, in the bottom-up stage, we firstly group attribute combinations into different clusters, each of which contains those attribute combinations affected by the same root cause only, with a robust probabilistic clustering method based on GRE.
In this way, \ours{} firstly breaks down the problem into simpler sub-problems (\ie{}, single root causes) and reduces search space.
Then in the top-down stage, we propose a score function, generalized potential score (GPS), to evaluate how likely a set of attribute combinations is the root cause, and search from each cluster the attribute combinations maximizing it with a efficient heuristic search strategy.
Finally, after the search, \ours{} determines external root causes based on GPS.

To evaluate \ours{}, we conduct extensive experimental studies based on two real-world datasets from two companies.
Since the real-world faults are not enough for evaluation, we propose a fault simulation method and obtain \num{5400} simulated faults.
The results show that \ours{} outperforms all baselines by 32.89\%  in different situations while keeping high efficiency (costs about 10s consistently for each fault).
We also inject 73 faults on an open-source benchmark system to prove the effectiveness of \ours{} in real-world scenarios.
For determining external root causes, the F1-score of \ours{} achieves 0.90 on average.
We also present several real-world success stories to demonstrate the efficacy of \ours{} in real-world systems.

The major contributions are summarized as follows:
\begin{itemize}
    \item We propose a novel property about root causes of multidimensional data, which is proved to hold in different situations and hold in real-world faults.
    \item We identify the importance of determining external root causes propose the first effective method for it.
    \item We propose a novel ``bottom-up\&top-down'' localization method, \ours{}, achieving high efficiency without much loss of genericness and robustness.
    \item We evaluate the effectiveness and efficiency \ours{} in different situations based on \num{5400} simulated faults and \num{XXX} injected faults.
    We make our dataset and implementation public to help further studies in the field\footnote{\url{https://github.com/NetManAIOps/PSqueeze}}.
\end{itemize}

This paper extends our previous conference paper, Squeeze \cite{li2019generic}, in four aspects.
\begin{itemize}
    \item New methods.
    First, this paper proposes the first external-root-cause-determining method in the field (\cref{sec:external-root-cause}).
    Second, to reduce the influence of noises, we propose a novel \textit{probabilistic} clustering method (\cref{sec:probabilistic-clustering}).
    Hence we name our new method as \ours{} (probabilistic Squeeze).
    \item New experiment settings.
    First, we propose a more reasonable fault simulation strategy for evaluation (\cref{sec:datasets}).
    Second, we implement and compare two more recent related works, MID~\cite{gu2020efficient} and ImpAPTr~\cite{rong2020locating}.
    Third, we introduce two new datasets based on fault injection on an open-source benchmark system.
    \item New experiment results based on the new experiment settings (\cref{sec:evaluation}) and new real-world success stories (\cref{sec:deployment}).
    The results show \ours{} is effective and efficient and outperform the previous approaches including Squeeze.
    \item Enhancement to presentation. 
    First, we clarify the definition of basic concepts formmaly (\cref{sec:background}).
    Second, we present more details about our methodology, such as the proof of GRE for productions (\cref{sec:generalized-ripple-effect}), and the reasons for our methodology's design choices (\cref{sec:methodology}).
    Notably, for a better understanding of GRE, we present a much more straightforward proof in \cref{sec:GRE-derived}.
\end{itemize}

%% file: background.tex
\section{Background}
\label{sec:background}

In this section, we first describe our problem intuitively.
Then we introduce some necessary concepts, notations, and definitions.
Finally, we define our problem formally.

\input{motivating_example}

\input{concepts_and_notations}

\input{problem_definition}

%% file: motivating_example.tex
\subsection{Root Cause Localization for Multi-Dimensional Data}
\label{sec:problem-introduction}
\begingroup
\setlength{\tabcolsep}{1pt} %
\renewcommand{\arraystretch}{1} %
\begin{table}[hbt]
\captionsetup{skip=0pt}
\centering
\caption{Example structured logs for an online shopping platform}
\footnotesize
\begin{tabular}{ccccc}
\toprule
Order ID & Timestamp & Dollar Amount & Province & ISP \\ \midrule
A001 & 2020.07.15 10:00:01 & \$16 & Beijing & China Mobile \\ 
A002 & 2020.07.15 10:00:05 & \$21 & Beijing & China Unicom \\
\bottomrule
\end{tabular}
\label{tbl:raw-log}
\end{table}
\endgroup

\begin{table}[hbt]
\captionsetup{skip=0pt}
\footnotesize
\begin{minipage}{1\columnwidth}
\caption{An example multi-dimensional data at a specific time point.}
\centering
\begin{tabular}{llll}
\toprule
Province                                & ISP                                    & real value                              & forecast value                            \\ \midrule
{\textbf{Beijing}} & {\textbf{China Mobile}} & {\textbf{5}} & {\textbf{10}} \\ 
{\textbf{Beijing}} & {\textbf{China Unicom}} & {\textbf{10}} & {\textbf{20}} \\ 
Shanghai                                & China Unicom                                 & 30                              & 31                               \\ 
Guangdong                               & China Mobile                                 & 10                                & 9.8                                 \\ 
Zhejiang                               & China Unicom                                 & 2                               & 2                                \\ 
Guangdong                               & China Unicom                                 & 200                               & 210                                \\ 
Shanxi                               & China Unicom                                 & 20                               & 22                                \\ 
Jiangsu                               & China Unicom                                 & 200                               & 203                                \\ 
Tianjin                               & China Mobile                                 & 41                               & 43                                \\ \midrule
\multicolumn{2}{c}{Total} & 518 & 550.8\\ \bottomrule
\end{tabular}
\label{tbl:ps-example}
\end{minipage}
\end{table}

As introduced in \cref{sec:introduction}, multi-dimensional data are essentially a group of structured logs generated by an online service system.
Specifically speaking, by grouping the logs (\eg{}, \cref{tbl:raw-log}) by some \textit{attributes} (\ie{}, \texttt{Timestamp}, \texttt{Province} and \texttt{ISP}) and aggregating the \textit{measure} values (\ie{}, \texttt{Dollar Amount}), we transform the original logs into multi-dimensional data (\eg{}, \cref{tbl:ps-example})~\cite{lin2020fast}.

To ensure service quality, operators closely monitor the overall measure values (\eg{}, total dollar amount).
When a measure value becomes abnormal (\eg{}, in \cref{tbl:ps-example}, the total dollar amount decreases from 550.8 to 518), a fault occurs in the online service system.
A fault usually causes only the measure values under specific attribute combinations abnormal in practice~\cite{gu2020efficient}.
For example, in \cref{tbl:ps-example}, when a fault happens in the servers in Beijing, only the measure values of (Province=Beijing) (\ie{}, the first two rows) are abnormal.
Such attribute combinations indicate the scope of the fault, and thus, are substantial clues to the underlying root causes.
We call such attribute combinations \textit{root-cause attribute combinations} and the set of root-cause attribute combinations as the \textit{root causes for the multi-dimensional data}~\cite{sun2018hotspot,lin2020fast,li2019generic}.
For convenience, without other conflicts, ``root causes'' in this paper refer to root causes for multi-dimensional data.
Localizing root causes of multi-dimensional data enables rapid fault diagnosis by directing the investigation.

\begin{table}[hbt]
\captionsetup{skip=0pt}
    \footnotesize
    \begin{minipage}{1\columnwidth}
    \caption{External root cause example.}
    \centering
    \begin{tabular}{lll}
    \toprule
    ISP                                    & real value                              & forecast value                            \\ \midrule
    {\textbf{China Mobile}} & {\textbf{56}} & {\textbf{62.8}} \\
    {\textbf{China Unicom}} & {\textbf{462}} & {\textbf{488}} \\ \midrule
    \multicolumn{1}{c}{Total} & 518 & 550.8\\ \bottomrule
    \end{tabular}
    \label{tbl:ExRC-example}
    \end{minipage}
\end{table}

The exact root causes (\eg{}, \{(Province=Beijing)\} in \cref{tbl:ExRC-example}) may contain uncollected attributes (\eg{}, \texttt{Province} in \cref{tbl:ExRC-example}).
We call such root causes \textit{external root causes}, which are not rare in practice~\cite{kim2013root}.
On the one hand, many attributes are seldom used in fault diagnosis because they contain many null values, are hard to understand, or are not informative. 
On the other hand, since the search space grows exponentially with the number of attributes, operators have to choose the most useful ones for root cause localization.
When external root causes exist, the localization results would be wrong and misleading.
However, all existing approaches do not check external root causes to our best knowledge.

%% file: concepts_and_notations.tex
\subsection{Necessary Concepts and Notations}
\label{sec:concepts-notations}

We denote the set of all \textit{attributes} of the studied multi-dimensional data $D$ as $A{=}\{a_1, a_2, ..., a_n\}$, where $a_i$ is the $i$-th attribute, and $n$ is the total number of attributes.
Each attribute has a finite number of discrete feasible values, which are called \textit{attribute values}.
We denote the set of attribute values of $a_i$ as $V_i{=}\{v_i^{(1)}{,}v_i^{(2)}{,}{...}{,}v_i^{(m_i)}\}$, where $v_i^{(j)}$ is the $j$-th attribute value of $a_i$, and $m_i$ is the number of attribute values of $a_i$.
An attribute $a_i$ and one of its attribute values $v_i^{(j)}$ construct a \textit{tuple}, $t=(a_i, v_i^{(j)})$.
For each attribute $a_i$, we denote the set of its tuples as $T_i=\{a_i\} \times V_i$, where ``$\times$'' refers to Cartesian product.
We denote the set of all tuples as $T=\cup_{i=1}^{n}T_i$.
Then an \textit{attribute combination} is a subset of $T$ that contains at most one tuple from each $T_i$.
Therefore \textit{the set of all attribute combinations} can be denoted as 
$E{=}\{e{\in}\mathcal{P}(T)\>|\>\forall T_i, |e\cap T_i|{\le} 1\}$, where $\mathcal{P}(T)$ refers to the power set of $T$, and $|\cdot|$ refers to the cardinality of a set.
In practice, a root cause of a fault can contain multiple root-cause attribute combinations, and a fault can have multiple root causes either. 
Hence \textit{the set of all root cause candidates} is $\mathcal{P}(E)$ rather than $E$.

A \textit{leaf attribute combination} $e$ (a.k.a. \textit{leaf} for simplicity) is an attribute combination that contains tuples of every attribute, \ie{}, $\forall T_i, |e{\cap} T_i|{=}1$.
An attribute combination $e_1$ is \textit{descended} from $e_2$ when $e_2{\subsetneq} e_1$.
For example, (Province=Shanghai$\land$ISP=China Unicom) is descended from (Province=Shanghai).
The insight is that if $e_1$ is descended from $e_2$, then the slice of data represented by $e_1$ is a subset of that of $e_2$.
We denote the set of all leaf attribute combinations descended from $e$ as 
$
    LE(e){=}\{e'{\in} E|e\subsetneq e'\>\land\>\forall T_i, |e'\cap T_i|{=}1\}
$.

\begin{figure}[htb]
\captionsetup{skip=0pt}
\centering
   \includegraphics[width=0.8\columnwidth]{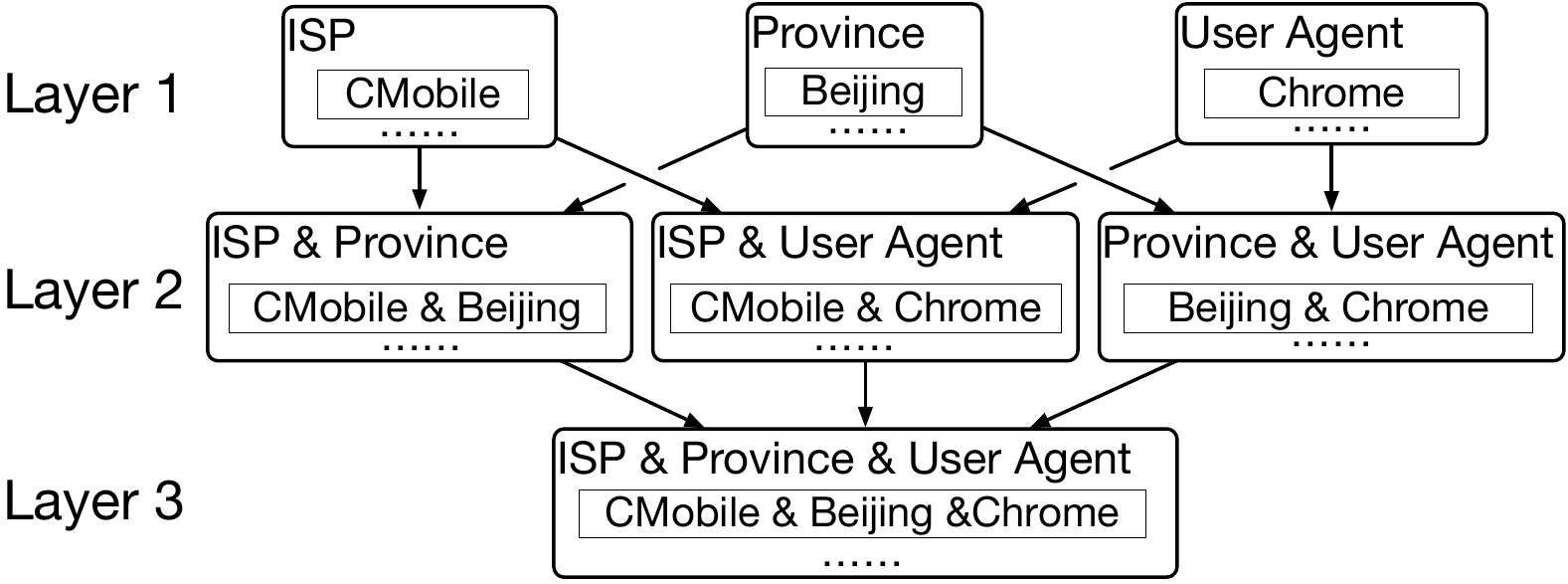}
   \caption{A graph of cuboids (rounded boxes) with 3 attributes}
  \label{fig:cuboid-graph}
\end{figure}

A \textit{cuboid} is a set of attribute combinations enumerating all attribute values for the involved attributes, as shown in \cref{fig:cuboid-graph}.
Given a set of attributes $A' \subset A$, the corresponding cuboid is 
$
Cuboid_{A'}=\{e{\in} E\>|\>\forall a_i \in A', |e\cap T_i|{=}1\>{\land}\> |e|{=}|A'|\}
$.
We call $|A'|$ the \textit{layer} of $Cuboid_{A'}$ (see \cref{fig:cuboid-graph}).

\textit{Fundamental measures} are those measures directly aggregated from raw logs and are additive~\cite{bhagwan2014adtributor}.
For example, in \cref{tbl:ps-example}, the total dollar amount of (Province=Beijing) is the sum of those of (Province=Beijing$\land$ISP=China Mobile) and (Province=Beijing$\land$ISP=China Unicom).
\textit{Derived measures} are derived from fundamental measures~\cite{bhagwan2014adtributor} and are typically non-additive~\cite{bhagwan2014adtributor}.
For example, the overall average dollar amount is not the sum over those of all ISPs.
The overall average dollar amount can be either greater than or less than that of (Province=Beijing).
Some previous approaches~\cite{lin2016idice,sun2018hotspot,gu2020efficient} are not applicable on derived measures due to these characteristics.

The \textit{real value} (denoted as $v(e)$) of an attribute combination $e$ is the measure value that is actually observed based on the raw transaction logs, and the \textit{forecast value} ($f(e)$) is its expected normal value.
We calculate the forecast value by a time-series forecasting algorithm (see \cref{sec:method-overview}).
Without loss of genericness, we assume both $v$ and $f$ are non-negative since most common measures are non-negative.
Note that for a derived measure $M=h(M_1, M_2, ..., M_l)$, $v_{M}(e)=h(v_{M_1}(e), v_{M_2}(e), ..., v_{M_l}(e))$.
Furthermore, for convenience, we extend the notations, $v$ and $f$, to sets of attribute combinations:
supposing that $S$ is a set of attribute combinations, for fundamental measures, $v(S){=}\sum_{e'\in \bigcup_{e\in S}LE(e)}v(e')$ and so does $f$.
For a derived measure $M=h(M_1, M_2, ..., M_l)$, $v_{M}(S)=h(v_{M_1}(S), v_{M_2}(S), ..., v_{M_l}(S))$ and so does $f$.

%% file: problem_definition.tex
\subsection{Problem Definition}
\label{sec:problem-definition}
The input of our problem is a snapshot of multi-dimensional data $D$ (with both real and forecasting values) at the fault time.
The forecasting values are obtained by an appropriate time-series forecasting algorithm, which is out of the scope of this paper.
To better evaluate the robustness of \ours{}, in this paper, we use MA (moving average, one of the simplest algorithms) for all scenarios (see later in \cref{sec:experiment-settings}).
Our goal is to localize the \textit{root cause of multi-dimensional data}, which refers to a set of attribute combinations that is:
\begin{itemize}
    \item \textit{Expressive}. 
    An expressive root cause candidate $S$ indicates the scope of faults in the multi-dimensional data accurately. 
    In other words, the part of $D$ specified by $S$ is abnormal, and the other part is normal.
    \item \textit{Interpretable}.
    An interpretable root cause candidate $S$ is as concise as possible to make operators focus on the faulty attributes and attribute values.
\end{itemize}
Sometimes there are multiple root causes at the same time, which indicate different underlying failures and have different influence.
In such cases, each root cause is a set of attribute combinations that is expressive and interpretable excluding the influence of the other root causes, and we aim to find the union of these root causes.
Following existing works~\cite{bhagwan2014adtributor,lin2016idice,sun2018hotspot,ahmed2017detecting,persson2018anomaly,gu2020efficient,lin2020fast,rong2020locating}, causal inference is also out of scope.

%% file: gre.tex
\section{Generalized Ripple Effect}
\label{sec:generalized-ripple-effect}

\subsection{Background of Ripple Effect}

\textit{Ripple effect}, first empirically observed by \cite{sun2018hotspot} for \textit{fundamental measures} only, captures the relationship of attribute combinations' abnormal magnitudes caused by the same root cause. 
The intuition is that all attribute combinations affected by the same root cause will change by the same proportion.
We denote the set of attribute combinations affected by a root cause $S \in \mathcal{P}(E)$ as $\text{Aff}(S)=\{e\in E\>|\>\exists e_0\in S,\>s.t.\>e_0\subseteq e\}$.
Then ripple effect can be expressed as
\begin{equation}
	{(f(e)-v(e))}/{f(e)}{=}{(f(S)-v(S))}/{f(S)}, \forall e {\in} \text{Aff}(S)
\label{eq:ripple-effect}
\end{equation}
For example, in \cref{tbl:ps-example}, the root cause is $S=\{(\textrm{Province}=\textrm{Beijing})\}$ in cuboid $C_{\textrm{province}}$.
Therefore, if $e_1=(\textrm{Province}=\textrm{Beijing}\land\textrm{ISP}=\textrm{China Unicom})$, then 
${(f(S)-v(S))}/{f(S)}={((10{+}20)-(5{+}10))}/{(10{+}20)}=0.5$ and 
${(f(e_1) - v(e_1))}/{f(e_1)}={(20-10)}/{20}=0.5$.
Note that for multiple-root-cause faults, $S$ only denotes a single root cause rather than the union of multiple root causes.

\subsection{Generalizing Ripple Effect for Derived Measures}
\label{sec:GRE-derived}
First, we generalize ripple effect to derived measures.
We aim to prove that \cref{eq:ripple-effect} holds for a derived measure when \cref{eq:ripple-effect} holds for all its underlying fundamental measures. 
Since most common derived measures are the quotient or product of two fundamental measures (\eg{}, average dollar amount and success rate), without much loss of genericness, we provide the proof for such derived measures only.
Compared with our previous conference version, we simplify the equations for clarity.

\subsubsection{Quotient}
Consider three measures, $M_1, M_2, M_3$, where $M_1$ and $M_2$ are fundamental measures and $M_3{=}{M_1}/{M_2}$.
Because $M_1$ and $M_2$ are fundamental measures and they follow \cref{eq:ripple-effect}, for both $M_i$ ($i=1,2$), 
${v_{M_i}(e)}/{f_{M_i}(e)}={v_{M_i}(S)}/{f_{M_i}(S)}$.
Therefore, 
\begin{equation}
\begin{aligned}
&\frac{f_{M_3}(S)-v_{M_3}(S)}{f_{M_3}(S)}=(\frac{f_{M_1}(S)}{f_{M_2}(S)}-\frac{v_{M_1}(S)}{v_{M_2}(S)})\frac{f_{M_2}(S)}{f_{M_1}(S)} \\
&=1-\frac{v_{M_1}(S)}{f_{M_1}(S)}\frac{f_{M_2}(S)}{v_{M_2}(S)}
=1-\frac{v_{M_1}(e)}{f_{M_1}(e)}\frac{f_{M_2}(e)}{v_{M_2}(e)}\\
&=1-\frac{v_{M_3}(e)}{f_{M_3}(e)}
=\frac{f_{M_3}(e)-v_{M_3}(e)}{f_{M_3}(e)}
\end{aligned}
\end{equation}

\subsubsection{Product}
Similarly, if $M_1$ and $M_2$ are fundamental measures and $M_3{=}{M_1}\cdot{M_2}$, then for both $M_i$ ($i=1,2$), 
${v_{M_i}(e)}/{f_{M_i}(e)}={v_{M_i}(S)}/{f_{M_i}(S)}$.
Therefore,
\begin{equation}
    \begin{aligned}
        &\frac{f_{M_3}(S)-v_{M_3}(S)}{f_{M_3}(S)}
        =\frac{f_{M_1}(S)f_{M_2}(S)-v_{M_1}(S)v_{M_2}(S)}{f_{M_1}(S)f_{M_2}(S)}\\
        &=1- \frac{v_{M_1}(S)}{f_{M_1}(S)}\frac{v_{M_2}(S)}{f_{M_2}(S)}
        =1 -\frac{v_{M_1}(e)}{f_{M_1}(e)}\frac{v_{M_2}(e)}{f_{M_2}(e)}\\
        &=1-\frac{v_{M_3}(e)}{f_{M_3}(e)}
        =\frac{f_{M_3}(e)-v_{M_3}(e)}{f_{M_3}(e)}                        
    \end{aligned}
\end{equation}

The core idea of our proof is \textit{finite difference}~\cite{jordan1965calculus}, and a similar method can be applied when dealing with other types of derived measures.
Though our proof technique can hardly be applied on specific types of derived measures (\eg{}, tail latency), our results can already cover most common measures (\eg{}, the four golden signals from Google SRE~\cite{murphy2016site}), including both fundamental measures and derived measures.

\subsection{Generalizing Ripple Effect for Zero Forecast Values}
\label{sec:GRE-zero-forecast}
\Cref{eq:ripple-effect} does not work for zero forecast values (\ie{}, $f(S)=0$) and thus is not robust enough.
To tackle it, we replace $f$ with $\frac{f+v}{2}$.
The intuition is to use $\frac{f+v}{2}$ to estimate $f$.
Therefore, the formulation of GRE is
\begin{equation}
	\frac{f(e)-v(e)}{f(e)+v(e)}=\frac{f(S)-v(S)}{f(S)+v(S)}, \forall e \in \text{Aff}(S)
\label{eq:generalized-ripple-effect}
\end{equation}
\Cref{eq:generalized-ripple-effect} is consistent with \cref{eq:ripple-effect}:
\begin{itemize}[leftmargin=1em]
\item If $f(e)=v(e)=0$ or $f(S)=v(S)=0$, the relationship between $e$ and $S$ is meaningless since there is actually no data for $e$ or $S$. In other situations, \cref{eq:generalized-ripple-effect} is always meaningful.
\item If $f(e)\neq 0$ and $f(S)\neq 0$, it is obvious that \cref{eq:ripple-effect} is equivalent to \cref{eq:generalized-ripple-effect}.
\item If $f(e)=0\neq v(e)$ or $f(S)=0\neq v(S)$, then \cref{eq:ripple-effect} does not hold. We extend the idea of ripple effect to more generic cases by modifying the calculation of anomaly magnitudes.
\end{itemize}

\subsection{Deviation Score and Expected Abnormal Value}
\label{sec:deviation-score-expected-abnormal-value}
In this paper, we utilize GRE by \textit{deviation scores} and \textit{expected abnormal values}.
According to \cref{eq:generalized-ripple-effect}, for any attribute combination $e$ that is affected by the same root cause $S$, the value of $\frac{f(e)-v(e)}{f(e)+v(e)}$ keeps invariant.
We define it as \textit{deviation score} of $e$ (denoted as $d(e)$).
According to GRE,  
\begin{equation}
\forall e \in \text{Aff}(S), d(e)={\big(f(S)-v(S)\big)}/{\big(f(S)+v(S)\big)}
\label{eq:deviation-score-value}
\end{equation}

Therefore, given a root cause candidate $S$ and any attribute combination $e \in \text{Aff}(S)$, if $S$ is the correct root cause, then ${d}(e){=}\frac{f(S)-v(S)}{f(S)+v(S)}$.
As a result, the \textit{expected abnormal value} of $e$ should be 
\begin{equation}
a(e)=f(e){\big(1-{d}(e)\big)}/{\big(1+{d}(e)\big)}
\label{eq:expected-abnormal-value}
\end{equation}
If $a(e)$ differs from $v(e)$ a lot, then the candidate $S$ breaks GRE and is not the correct root cause.

%% file: methodology.tex
\section{Methodology}
\label{sec:methodology}

\subsection{Overview}
\label{sec:method-overview}

\begin{figure}[htb]
\captionsetup{skip=0pt}
\centering
  \includegraphics[width=\columnwidth]{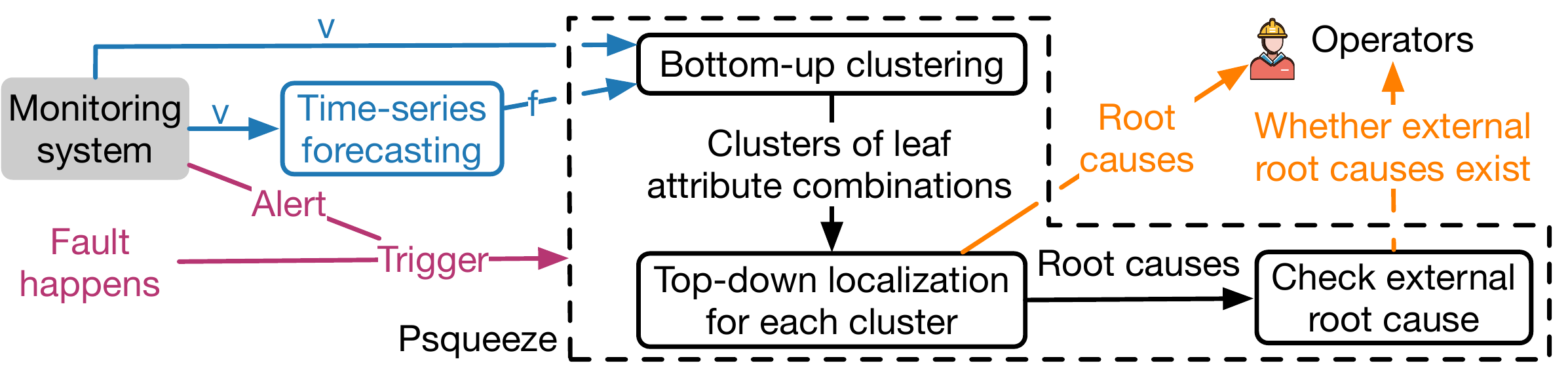}
  \caption{The workflow of \ours{}}
\label{fig:workflow}
\end{figure}

The workflow of \ours{} is illustrated in \cref{fig:workflow}, where the dashed box highlights the scope of \ours{}.
When a fault happens (often indicated by alerts from the monitoring system), \ours{} is triggered.
\ours{} takes the corresponding multi-dimensional data at the fault time ($v$) and its forecast values ($f$) as inputs.
Then, \ours{} reports the root causes to operators, and notifies operators whether there can be external root causes to avoid misleading.

\ours{} contains three stages: 1) bottom-up clustering (\cref{sec:bottom-up}), 2) top-down localization for each cluster (\cref{sec:top-down}), and 3) external root cause determining (\cref{sec:external-root-cause}).
Different from all previous work~\cite{bhagwan2014adtributor,persson2018anomaly,lin2016idice,sun2018hotspot,ahmed2017detecting,gu2020efficient,lin2020fast}, \ours{} employs a novel ``bottom-up then top-down'' searching strategy.
In the bottom-up stage, \ours{} groups leaf attribute combinations into different clusters, each of which contains the leaf attribute combinations affected by the same root cause.
The bottom-up clustering enables the further design of our efficient in-cluster localization method by simplifying the problem from multiple-root-cause localization to single-root-cause localization.
In the top-down step, \ours{} uses a heuristic method based on our proposed generalized potential score (GPS) to efficiently search for the root cause in each cluster output by the bottom-up step.
At the final stage, \ours{} determines whether there are external root causes.

Notably, this paper extends our previous conference version with respect to methodology in two aspects.
First, we employ probabilistic clustering for robustness (\cref{sec:probabilistic-clustering}).
Second, we introduce external root causes checking to avoid misleading (\cref{sec:external-root-cause}).
Besides, we fix the issue in \cref{eq:trace-off-weight} and enhance the presentation of the methodology for clarity.

\input{method/bottom_up}

\input{method/top_down}

\input{method/ExRC}

%% file: method/bottom_up.tex
\subsection{Bottom-Up Searching through Clustering}
\label{sec:bottom-up}
\label{sec:probabilistic-clustering}

In this stage, we determine the cluster boundaries with a probabilistic cluster method based on the leaf attribute combinations with large abnormal changes.

\subsubsection{Forecast Residual-Based Filtering}
\label{sec:forecast-residual-filtering}
\begin{figure}[htb]
\captionsetup{skip=0pt}
\centering
  \includegraphics[width=0.5\columnwidth]{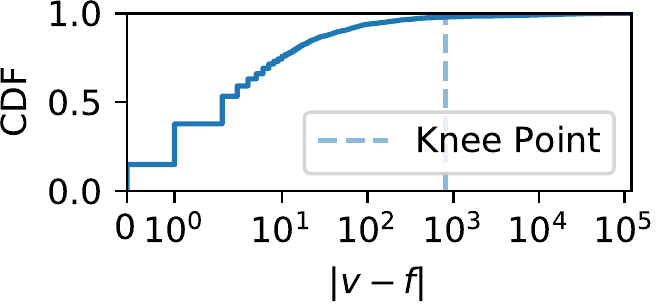}
  \caption{An illustration example of our forecast residual-based filtering}
\label{fig:forecast-residual-filtering}
\end{figure}

In order to make the following clustering step focus on abnormal leaf attribute combinations (\ie{}, affected by root causes), we need to detect abnormal leaf attribute combinations.
Following existing work~\cite{xu2018unsupervised,liu2015opprentice,li2018robust,sun2018hotspot}, we use \textbf{forecast residuals} (\ie{}, the difference between real and forecast values) to indicate the extent of the changes of attribute combinations and apply a threshold to decide whether the change is abnormal or not.
We apply knee-point method on the cumulative distribution function (CDF) of the forecast residuals of leaf attribute combinations for automated threshold selection.
It is because given a large number of leaf attribute combinations, the number of abnormal leaf attribute combinations is usually much less than the normal ones. 
A knee point refers to a point where the increase of filtered-out leaf attribute combinations is no longer worth the increase of the threshold.
In \cref{fig:forecast-residual-filtering}, we present an example CDF of an online service system fault and its knee point (the vertical dashed line).
We define a knee point as the point with maximum curvature rather than other ad-hoc definitions for genericness and robustness following existing work~\cite{satopaa2011finding}.
The advantage of the knee-point method is that it is simple, efﬁcient, and completely automated.

\subsubsection{Calculating the Distribution of Deviation Scores}
\label{sec:distribution-deviation-scores}

\begin{figure}[hbt]
\captionsetup{skip=0pt}
\centering
    \includegraphics[width=0.9\columnwidth]{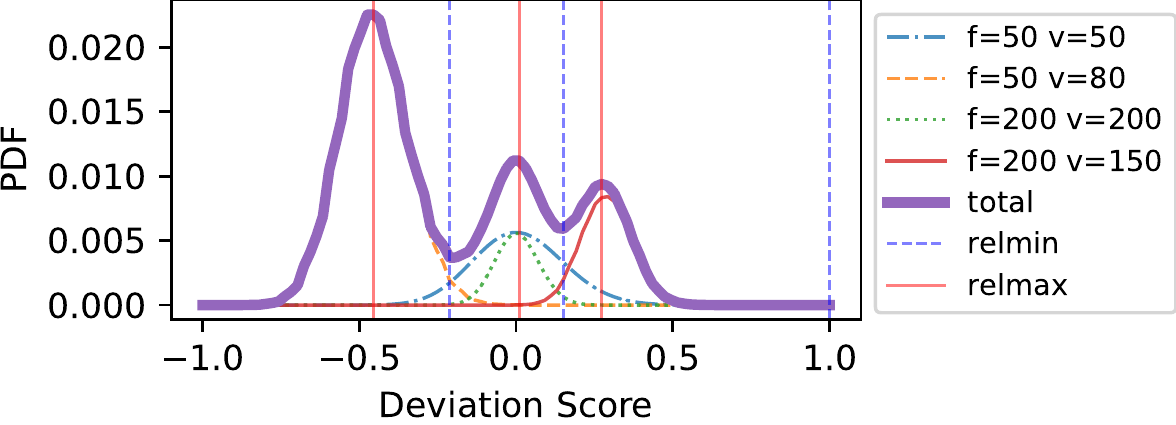}
    \caption{Illustration of \ours{}'s probabilistic clustering.}
    \label{fig:illustrate-probabilistic-cluster}
\end{figure}

As introduced in \cref{sec:deviation-score-expected-abnormal-value}, leaf attribute combinations affected by the same root causes have the same deviation scores, based on which we design our clustering method.
First, we estimate the distribution of all leaf attribute combinations' deviation scores.
Squeeze (our previous conference version~\cite{li2019generic}) presumes the observed deviation score is correct.
However, deviation scores can be significantly affected by noises from natural variation or inaccurate forecasting, especially when the real and forecast values are small.
For example, given an attribute combination where $v\sim Pois(5)$ and $f=5$, its deviation score is supposed to be $0$ since the expectation of its real value is equal to its forecast value.
However, its deviation score could be $\pm 0.222$ or $\pm 0.182$ ($v=4,6$ due to noise or natural variation, or $f=4,6$ due to inaccurate forecast), and thus it can be mistakenly grouped into incorrect clusters.
Thus it is required to estimate the distribution of deviation scores in a more robust manner.
Since the variation and forecasting errors can hardly be eliminated, we try to explicitly model the noises.
Unlike Squeeze, \ours{} considers what if the deviation scores of an attribute combination are biased and determines the probability that the leaf attribute combination should be grouped into each cluster.
As a result, when the deviation score of an attribute combination is largely biased due to noises and thus is grouped into an incorrect cluster by Squeeze, \ours{} groups it into the correct cluster with a certain probability.
In this way, we make \ours{} more robust to such noises than Squeeze.

Specifically speaking, we firstly calculate the probability density function (PDF/PMF) for each abnormal attribute combination's deviation score and then average them to obtain the overall distribution.
The choice of PDF and PMF depends on our measure: we calculate PDF for continuous measures (\eg{}, average response time) and PMF for discrete measures (\eg{}, the number of orders).
To calculate the PDF/PMF, we need to assume the distribution family according to domain knowledge about the nature of the measure.
For example, considering the measure of the number of total orders, we can assume that the number of total orders $v$ follows \textit{Poisson distribution}, \ie{}, $v\sim Pois(\lambda=v')$, because the speed of order arrivals keeps stable in a short duration.
Then the real deviation score is supposed to be $ds=2\frac{f-v'}{f+v'}$.
Therefore, the probability mass function of $ds$ is $P(ds=2\frac{f-v-k}{f+v+k})=Pois(v;\lambda=v+k)$ where $v$ denotes the observed real value.
Compared with that Squeeze considers that $ds$ follows $P(ds=2\frac{f-v-k}{f+v+k})=\begin{cases}1,  & k=0 \\ 0 & k \neq 0\end{cases}$, it models the probability that the observed deviation score is biased due to noises.
In this way, we calculate all the PDF/PMFs of leaf attribute combinations' deviation scores (\eg{}, the light curves in \cref{fig:illustrate-probabilistic-cluster}).
Then, by averaging all these probability density/mass functions together, we get the overall distribution of deviation scores of all attribute combinations (\eg{}, the bold solid curve in \cref{fig:illustrate-probabilistic-cluster}).

In this paper, we use \textit{Poisson distribution} for all fundamental measures, including \#orders and \#page views.
It is because \textit{Poisson distribution} is suitable for describing the number of event occurrences with a constant mean rate.
For derived measures, we do not use probabilistic clustering, given the difficulty in finding appropriate distribution families for derived measures.

\subsubsection{Determining the Cluster Boundaries}
\label{sec:cluster-boundaries}

 \begin{algorithm}
    \caption{Deviation Score Based Probabilistic Clustering}
    \begin{algorithmic}[1]
        \Procedure{DensityCluster}{$PDF$}
        \State $centers \leftarrow \text{argrelmax}(PDF)$ %
        \State $boundaries \leftarrow \text{argrelmin}(PDF)$ %
        \State $clusters \leftarrow []$
        \For{$center$ in $centers$}
        \State $l\leftarrow$ maximum in $boundaries$ s.t. $l<center$
        \State $r\leftarrow$ minimum in $boundaries$ s.t. $r>center$
        \State $clusters$.append($\{e \in LE(\emptyset)|d(e)\in [l, r]\}$)
        \EndFor
        \Return clusters
        \EndProcedure
    \end{algorithmic}
    \label{ago:cluster}
\end{algorithm}

Based on the distribution of deviation scores, we determine the number of clusters and the cluster boundaries.
Since the deviation scores of the leaf attribute combinations affected by the same root cause should crowd around a small area, the low-density areas separate the clusters.
It is a 1-dimensional clustering problem, which is special because there is no saddle point in 1-dimensional spaces, and thus \textit{relative maximums} of the PDF/PMF represent all high-density areas.
Hence applying a high-dimensional clustering method (\eg{}, DBSCAN~\cite{schubert2017dbscan}) would be unnecessarily costive.
Instead, we design a simple clustering method intuitively, as shown in \cref{ago:cluster}.
First, we select the points where deviation scores crowd, \ie{}, relative maximums of the PDF/PMF (\eg{}, the three solid red vertical lines in \cref{fig:illustrate-probabilistic-cluster}), as the centroids of clusters.
Hence the number of clusters is equal to the number of relative maximums of the PDF/PMF.
Then, since low-density areas separate the clusters, we select the nearest low-density points, \ie{}, \textit{relative minimums} of the PDF (\eg{}, the dashed purple vertical lines in \cref{fig:illustrate-probabilistic-cluster}), as the boundaries of the clusters.
Then, the probability of an attribute combination in a cluster is equal to the probability that its deviation score locates between the boundaries.
In other words, a cluster contains the leaf attribute combinations whose PDF/PMF intersects with the area between boundaries, and the intersection area denotes the probability that the attribute combination is in the cluster.

%% file: method/top_down.tex
\subsection{Top-Down Localization in Each Cluster}
\label{sec:top-down}
\label[section]{sec:localization}

\begin{algorithm}[hbt]
    \caption{Localization in Each Cluster}
    \begin{algorithmic}[1]
        \Procedure{InClusterLocalization}{$cluster$}
        \State $root\_causes \leftarrow []$
        \For{$cuboid$ in all cuboids from top to bottom} \label{alg:line:cuboid-wise-search}
        \State $AC$ $\leftarrow$ sorted attribute combinations in $cuboid$ by $r_{descended}$ in descending order \label{alg:line:descended-ratio}
        \For{$split$ in all valid splits}
        \State $score[split]\leftarrow \text{GPS}(AC[:split])$
        \EndFor
        \State $root\_cause\leftarrow AC[:\text{argmax}_{split}score]$ \label{alg:line:max-GPS}
        \State $root\_causes \leftarrow root\_causes + [root\_cause]$
        \If{$root\_cause$'s $score$ $\ge$ $\delta$} \label{alg:line:stop-search-next-layer}
        	\State Stop search next layer
        \EndIf
        \EndFor
        \State sort $root\_causes:=\{S_i\}$ by $GPS(S_i) * C - I(S_i)$ in descending order\label{alg:line:selection-from-cuboids}
        \State \Return $root\_causes[0]$
        \EndProcedure
    \end{algorithmic}
    \label[algorithm]{ago:localization}
\end{algorithm}

The output of the bottom-up search is a list of clusters, each of which is a set of leaf attribute combinations that are affected by the same root cause.
For convenience, we denote a cluster as $Cluster$. 
At this stage, we aim to localize root causes for each cluster.
The root cause of a cluster is defined as a set of attribute combinations that is expressive when considering $Cluster$ and the normal leaf attribute combinations only and is interpretable.
To overcome the challenges of huge search space, we propose an efficient heuristic search method for the in-cluster root cause localization, which contains three key techniques.
1) a cuboid-wise top-down search strategy to narrow down the search space,
2) a heuristic strategy to search a cuboid efficiently,
and 3) a robust objective function to evaluate the expressiveness and interpretability of a root cause candidate.
A summary of our method at this stage is presented in \cref{ago:localization}.

\subsubsection{Cuboid-Wise Search Strategy}
We search for each cluster's root cause in each cuboid layer by layer (\cref{alg:line:cuboid-wise-search}).
Taking \cref{fig:cuboid-graph} as an example, we would search cuboid ISP, Province, and User Agent first, then search cuboid ISP\&Province, ISP\&User Agent, and Province\&User Agent, and finally search cuboid ISP \& Province\&User Agent.
On the one hand, the cuboid-wise search strategy is motivated by the assumption that the root cause of a cluster is only a subset of a cuboid.
The assumption is practical due to the following intuitions: 
1) the attribute combinations in the cluster are affected by the same root cause;
2) in practice one root cause rarely requires more than one cuboid to describe it according to our analysis on many real-world production faults.
On the other hand, we search shallow cuboids ﬁrst because root cause candidates in shallower cuboids are more interpretable (see \cref{sec:problem-definition}) than those in deeper cuboids, and thus we call it a top-down search method.

\subsubsection{Heuristic Method to Search a Cuboid}
If an attribute combination $e$ is part of the root cause, then according to GRE, all of its descent leaf attribute combinations (\ie{}, $LE(e)$) should have similar deviation scores, \ie{}, they should all in the cluster $Cluster$.
We call the ratio of descended leaf attribute combinations in the cluster of $e$ as the \textit{descended ratio} of $e$.
It is denoted as 
\begin{equation}
    r_{descended}(e)=\frac{\sum_{e'\in LE(e)\cap Cluster}p(e'\in Cluster)}{\sum_{e'\in LE(e)}p(e'\in Cluster)}
\end{equation}
, where $p(e'\in Cluster)$ denotes the probability that $Cluster$ contains $e'$ (see \cref{sec:cluster-boundaries}).
For example, in \cref{tbl:ps-example}, supposing that the cluster contains the first two rows in the table, and we are searching cuboid Province now, the descended ratio of $(Province=Beijing)$ is $1$ and those of others are $0$. 
We sort the attribute combinations of a cuboid by their descended ratios in descending order (\cref{alg:line:descended-ratio}).
In this way, the attribute combinations at the front of the list are more likely to be part of the root cause than those at the back.

\subsubsection{Generalized Potential Score}
\label{sec:generalized-potential-score}
Then, we aim to find the top-$k$ items in the sorted attribute combination list of a cuboid as the root cause candidate of the cuboid (\cref{alg:line:max-GPS}).
For this purpose, we propose \textit{generalized potential score} (GPS), to evaluate how likely a set of attribute combination ($S$) is the root cause in the following aspects:
1) it is expressive, \ie{}, the real and forecast values of its descended leaf attribute combinations (\ie{}, $LE(S)=\bigcup_{e\in S}LE(e)$) should be different, and those of other leaf attribute combinations should be close;
2) it follows GRE, \ie{}, the real values of $LE(S)$ should be close to the corresponding expected abnormal values (see \cref{sec:deviation-score-expected-abnormal-value}).
We do not evaluate interpretability here because in the same cuboid, adding extra attribute combinations reduce expressiveness and interpretability simultaneously.
Comparing real and expected abnormal values helps to filter the false expressive candidates caused by inaccurate forecasting and noise and make \ours{} more robust.

We measure the difference between the real and forecast values of $LE(S)$ by normalized Manhattan distance, \ie{}, $d_1(\vv{v}(S), \vv{f}(S))=\frac{1}{|LE(S)|}\sum_{e\in LE(S)}|v(e){-}f(e)|$, and the difference between the real and expected abnormal values by $d_1(\vv{v}(S), \vv{a}(S))=\frac{1}{|LE(S)|} \sum_{e\in LE(S)}|v(e)-a(e)|$.
Similarly, for other leaf attribute combinations (denoted as $LE(S')$), the difference between the real and forecast values is  $d_1(\vv{v}(S'), \vv{f}(S'))=\frac{1}{|LE(S')|}\sum_{e\in LE(S')}|v(e){-}f(e)|$.
Considering that we are conducting in-cluster localization and there can be other root causes, $LE(S')$ does not contain the leaf attribute combinations in other clusters.
Based on these, we define GPS as follows:
\begin{equation}
	GPS=1 - \frac{d_1(\vv{v}(S), \vv{a}(S))+d_1(\vv{v}(S'), \vv{f}(S'))}{d_1(\vv{v}(S), \vv{f}(S)))+d_1(\vv{v}(S'), \vv{f}(S'))}
\label{eq:localization-score}
\end{equation}

GPS robustly indicates $S$'s expressiveness, even if the anomaly magnitude of $S$ is insignificant.
Considering \textit{potential score}~\cite{sun2018hotspot} and \textit{explanation power}~\cite{bhagwan2014adtributor,persson2018anomaly}, the forecast residuals of $LE(S')$ would accumulate as the size of $LE(S')$ increases, and thus they cannot reflect the expressiveness when the anomaly magnitude of $S$ is insignificant.
For example, in \cref{tbl:ps-example}, the GPS of the first two rows in bold is $0.743$, while the potential score is $0.303$ and the explanation power is $0.457$.

\subsubsection{Selecting among Candidates from Different Cuboids}
Finally, we select the root cause with both high expressiveness and interpretability from the candidates found in the cuboids.
For this purpose, we quantitively define the interpretability of $S$ as $I(S)=\sum_{e\in S}|e|^2$.
For example, $I(\text{\{(Province=Beijing)\}})=1$ and $I(\text{\{(Province=Beijing}\land\text{ISP=CUnicom), (Province=Beijing}\land\text{ISP=CMobile)\}})=8$.
We use a weight $C$ to trade off expressiveness (measured by GPS) and interpretability (\cref{alg:line:selection-from-cuboids}).
To calculate $C$ automatically, we employ an empirical formula, which can achieve good effectiveness:
\begin{equation}
\begin{aligned}
    g_{cluster}&={\log(num\_cluster + 1)} / {num\_cluster} \\
	g_{attribute}&={num\_attr} / {\log(num\_attr + 1)} \\
	g_{coverage}&= -\log (\text{coverage of abnormal leaves}) \\
	C &= g_{cluster} \times g_{attribute} \times g_{coverage}
\end{aligned}
\label{eq:trace-off-weight}
\end{equation}
The intuition is that if there are fewer clusters or more attributes, or the cluster contains fewer abnormal leaf attribute combinations, then GPS is more important than interpretability.
It is refined based on that in our previous conference version~\cite{li2019generic}, and the original version can be negative in some cases and causes errors.

Furthermore, for efficiency, if the candidates' GPS scores exceed a given threshold $\delta$ at a certain layer of cuboids,  \ours{} would stop searching deeper layers (\cref{alg:line:stop-search-next-layer}).
We set $\delta=0.9$ by default and discuss its impact in \cref{sec:different-configurations}.

%% file: method/ExRC.tex
\subsection{Determine External Root Cause}
\label{sec:external-root-cause}
\begin{algorithm}[h]
    \caption{Determine External Root Cause}
    \begin{algorithmic}[1]
        \Procedure{determine\_ExRC}{$rc\_list$}
        \State $minGPS$ $\leftarrow +\infty$ 
        \For{$S$ in $rc\_list$}
        \State $minGPS \leftarrow \min(GPS(S), minGPS)$
        \EndFor{}
        \Return $minGPS\ge \delta_{ExRC}$
        \EndProcedure{}
    \end{algorithmic}
\label{alg:external-root-cause}
\end{algorithm}

\ours{} determines external root causes by examining whether the localized root causes are expressive enough, \ie{}, have high GPS scores.
When there are external root causes, since it is impossible for algorithms to localize the real root causes, the GPS scores of the localized root causes would be relatively low.
Specifically speaking, we check whether the GPS scores of the clusters is less than an automated threshold (denoted by $\delta_{ExRC}$).
We denote the minimum GPS scores of all per-cluster root cause as $min\_GPS$.
If $min\_GPS< \delta_{ExRC}$, there is at least one cluster where \ours{} is not able to find a good enough root cause, which indicates that there probably exist external root causes.
Then the operators will be informed that the results of \ours{} can be misleading due to external root causes.
If $min\_GPS\ge \delta_{ExRC}$, then \ours{} localizes good enough root causes for all selected clusters, then there are not external root causes, and thus the final result is reliable.

The threshold $\delta_{ExRC}$ is automatically selected by historical data.
The key idea is that for those faults without external root causes, their $min\_GPS$, which represents the minimum GPS of all per-cluster root causes, should be near 1.
It is because a good root cause should be localized for each cluster, and thus they would be grouped into a cluster by the density-based clustering method in \cref{sec:bottom-up}.
Therefore, given those $min\_GPS$ (defined in \cref{alg:external-root-cause}) of historical faults, we firstly cluster them with \cref{ago:cluster}, and then use the lower boundary of the cluster with the largest centroid as $\delta_{ExRC}$.
Note that we \textbf{do not} need to know which faults encounter external root causes, \ie{}, the automated threshold selection is unsupervised.
If there are not enough historical faults, we use a default value $0.8$ for $\delta_{ExRC}$.

%% file: evaluation.tex
\section{Evaluation}
\label{sec:evaluation}
In this section, we conduct extensive experiments to evaluate the localization accuracy and efficiency of \ours{} based on both simulated and injected faults.

\input{evaluation/experiment_settings}

\input{evaluation/overall_performance}
\input{evaluation/performance_exrc}
\input{evaluation/efficiency}
\input{evaluation/different_configuration}

%% file: evaluation/experiment_settings.tex
\subsection{Experiment Settings}
\label{sec:experiment-settings}

\subsubsection{Datasets}
\label{sec:datasets}
We have two real-world datasets collected from two production systems of two companies in several weeks.
One is from an online shopping platform (denoted as $\mathcal{I}_1$), and the other one is from an Internet company (denoted by $\mathcal{I}_2$).
Although these are real-world data, it is hard to obtain enough real-world anomalies and the corresponding root causes as the ground truth.
Therefore, we generate simulated faults according to GRE based on these real-world datasets to evaluate \ours{} as follows:
\begin{enumerate}
    \item We select a time point from the time series and add different Gaussian noises to the real values of all leaves.
    It is used to emulate different forecast residuals.
    \item We randomly choose $n\_element\in\{1,2,3\}$ (a.k.a. $n\_ele$ for short) cuboids \textit{with replacement} in layer $cuboid\_layer$, which is randomly chosen from $\{1,2,3\}$.
    \item We randomly choose $n\_element$ different attribute combinations from every selected cuboid, which are the root-cause attribute combinations for this simulated fault.
    Note that the datasets with only different forecast residuals (\ie{}, $\mathcal{B}_1, \mathcal{B}_2, \mathcal{B}_3, \mathcal{B}_4$) share the same time points but have different root-cause attribute combinations.
    \item For each selected root-cause attribute combination, we modify the real values of its descended leaf attribute combination by GRE with a random magnitude.
    Particularly, for $\mathcal{D}$, the measure of which is success rate, we firstly modify their success rates according to GRE.
    Then, we randomly generate the total order numbers and successful order numbers according to the success rates.
    \item We add extra Gaussian noises ($\mathcal{N}(0, 5\%)$) to these descended leaf attribute combinations since GRE would not perfectly hold in practical faults.
    \item We drop invalid ``faults'' when
    \begin{enumerate} 
    \item there exists another attribute combination that shares the same (or very similar) set of descended leaf attribute combinations with a selected root-cause attribute combinations; 
    \item the Gaussian noises added in normal leaf attribute combinations (\ie{}, those leaf attribute combinations that are not descended from any selected root-cause attribute combination) are so large that the overall measure value of them is abnormal.
    \end{enumerate}
\end{enumerate}

The root cause attribute combinations with different deviation scores are simulating independent multiple root causes.
Note that the deviation score of different root-cause attribute combinations could be the same sometimes.
In such cases, the root-cause attribute combinations with the same deviation score would be considered as one root cause containing multiple root-cause attribute combinations.
In all datasets, we add anomalies with random magnitudes, and thus, the anomalies are not guaranteed to be significant.

\begin{figure}[hbt]
\captionsetup{skip=0pt}
\centering
\includegraphics[width=0.7\columnwidth]{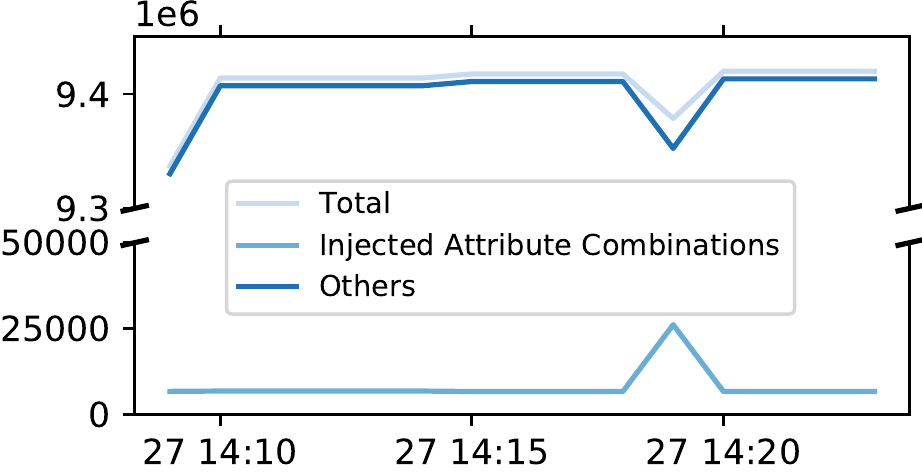}
\caption{The measure value of attribute combinations that are not intended to simulate faults on is significantly abnormal.}
\label{fig:old-dataset}    
\end{figure}

This paper extends the simulation process in our previous version~\cite{li2019generic} by adding step 6.
The process described in step 1$\sim$5 could generate inappropriate faults for evaluation.
For example, consider that we randomly choose (A=a1$\land$B=b1$\land$C=c1) as the root-cause attribute combination, and b1 is related to c1.
In such cases, (A=a1$\land$B=b1$\land$C=c1), (A=a1$\land$B=b1) and (A=a1$\land$C=c1) have almost the same set of leaf attribute combinations.
Therefore, it would be unreasonable to take (A=a1$\land$B=b1$\land$C=c1) as the root cause rather than (A=a1$\land$C=c1) or (A=a1$\land$B=b1).
Nevertheless, as we add Gaussian noises in all leaf attribute combinations, thus the attribute combinations that are not intended to simulate faults on would become abnormal as well, as shown in \cref{fig:old-dataset}.
To tackle this problem, unlike our previous version~\cite{li2019generic}, we remove invalid faults by step 6.

\begingroup
\setlength{\tabcolsep}{2pt} %
\renewcommand{\arraystretch}{1} %
\begin{table}[htb]
\captionsetup{skip=0pt}
\caption{Summary of Datasets}
\footnotesize
\centering
\begin{tabular}{ccccccccc}%
    \toprule%
    &$n$&$|LE(\emptyset)|$&$|\mathcal{P}(E)|$&\textbf{Source}&\textbf{Measure}&\textbf{Residual}&\textbf{EP}\\%
    \midrule%
    $\mathcal{A}$ &5&15324&$2^{75888}$&$\mathcal{I}_1$&F: \#orders&3.92\%&90.3\%\\%
    $\mathcal{B}_1$&4&21600&$2^{31338}$&$\mathcal{I}_2$&F: \#page views&3.97\%&74.2\%\\%
    $\mathcal{B}_2$&4&21600&$2^{31338}$&$\mathcal{I}_2$&F: \#page views&7.96\%&60.6\%\\%
    $\mathcal{B}_3$&4&21600&$2^{31338}$&$\mathcal{I}_2$&F: \#page views&11.9\%&56.1\%\\%
    $\mathcal{B}_4$&4&21600&$2^{31338}$&$\mathcal{I}_2$&F: \#page views&15.9\%&53.7\%\\%
    
    $\mathcal{D}$&4&13806&$2^{21534}$&$\mathcal{I}_2$&D: success rate&3.99\%&59.3\%\\%

    $\mathcal{E}$&9&373&$2^{373}$&-&D: average latency&37.3\%&89.5\%\\%
    $\mathcal{F}$&9&373&$2^{373}$&-&D: stall rate&45.8\%&86.9\%\\%
    
    \bottomrule%
    \end{tabular}%
\label{tbl:datasets}
\end{table}
\endgroup

\textbf{Simulated fault datasets.}
By the new simulation method, we get 6 new simulated fault datasets with different simulation parameters.
In \cref{tbl:datasets}, we describe some basic statistics of our datasets.
For each combination of $n\_ele$ and $cuboid\_layer$, we simulated \num{100} faults in each dataset.
Hence there are \num{5400} faults in total.
In \cref{tbl:datasets}, $n$ denotes the number of attributes, $|LE(\emptyset)|$ denotes the number of all leaf attribute combinations, and $|\mathcal{P}(E)|$ denotes the number of all root cause candidates.
These datasets contain three different measures, including both fundamental (denoted as $F$ in \cref{tbl:datasets}) and derived measures ($D$), and all of them are of common golden signals~\cite{murphy2016site}.
In practice, we do not count the attribute combinations that never occur in data, and thus $|LE(\emptyset)|$ and $|\mathcal{P}(E)|$ seem lower than theoretical results.
Residual in \cref{tbl:datasets} denotes the average forecasting residuals in percent of all normal leaf attribute combinations.
EP (explanation power~\cite{bhagwan2014adtributor,persson2018anomaly}) denotes the fraction of total forecast residual of all abnormal leaf attribute combinations over that of both normal and abnormal ones.

\textbf{Injected fault datasets.}
Furthermore, compared with the previous conference version, this paper also two new datasets, named \ds{E} and \ds{F}, based on more realistic fault injection (rather than directly adjusting measure values).
More specifically, we deploy Train-Ticket~\cite{zhou2018fault}, which is one of the largest open-source microservice benchmark systems and is widely used in literature~\cite{li2021practical,yu2021microrank,zhou2018fault,chen2022deep,li2022actionable}, on a Kubernetes cluster with five servers.
We utilize Istio~\cite{istio} and ChaosMesh~\cite{2022chaosmesh} to inject the following types of faults onto the system: delaying or dropping packets or HTTP requests/responses sent to specific pods/APIs/services or containing specific parameters.
We injected 73 faults in total.
Then, we collect detailed information on every HTTP request between the microservices with Jaeger and Istio, including client/server service name, URL, response time, status code, etc.
Based on the request details, we collect two types of derived measures to build the two datasets, i.e., average latency (=total latency of all requests / the number of requests) and stall rate (=the total number of stall requests/the number of requests).
In both datasets, the attributes and number of distinct attribute values are as follows: client service name (17), pod (64), method (3), URL prefix (71), station name (10), train type (6), start station (9), end station (10).
Compared with the simulated faults (i.e., \ds{A}, $\mathcal{B}_*$, and \ds{D}), the diversity of root causes is limited with respect to $n\_elements$ and $cuboid\_layer$, but the injected faults do not rely on any assumption and are much more representative of real-world system failures.

\subsubsection{Evaluation Metrics}
F1-score is used in this paper to evaluate root cause localization for multi-dimensional data.
We denote the root cause reported by the algorithm as $S$ and the ground truth as $\hat{S}$.
Then $tp{=}|S\cap \hat S|$ denotes the number of true positive root-cause attribute combinations, $fp{=}|S - \hat S|$ denotes the number of false positives, and $fn{=}|\hat S - S|$ denotes the number of false negatives.
Then F1-score is defined as:
\begin{equation}
F1\text{-}Score{=}{(2 \times tp)}/{(2 \times tp+ fp + fn})
\label{eq:f1-score}
\end{equation}
To extensively study the performance of \ours{} under various situations, following existing work~\cite{sun2018hotspot}, we evaluate F1-scores separately for different root cause settings, \ie{}, different $n\_elements$ and $cuboid\_layer$.

To evaluate \ours{} on determining external root causes, we also use the widely-used F1-score.
We denote the set of faults with external root causes as $\hat F$ and the set of faults that are reported by our algorithm to have external root causes as $F$. 
Then ExRC\_F1-score (\textbf{ex}ternal \textbf{r}oot \textbf{c}ause F1-score) is calculated as follows:
$$
ExRC\_F1\text{-}Score=2\times \frac{precision \times recall}{precision + recall}
$$
where $precision=|F\cap \hat F|/|F|$ denotes the probability that a determined external root cause is true and $recall=|F\cap \hat F|/|\hat F|$ denotes the fraction of external root cause cases that have been determined. 

Finally, we also evaluate the time efficiency of \ours{}.
In the following experiments, we present the average running time of all cases in the corresponding setting.

\subsubsection{Baseline Approaches}

We compare \ours{} with the following baseline approaches, which are summarized in \cref{tbl:related-works}:
\begin{itemize}
    \item Squeeze~\cite{li2019generic} (SQ), our previous conference version.
    \item Adtributor~\cite{bhagwan2014adtributor} (ADT) assumes root causes involve only single attributes and mines all attribute combinations with high \textit{explanation power} and then sorts them by \textit{surprise}.
    \item R-Adtributor~\cite{persson2018anomaly} (RAD) recursively calls Adtributor to localize multi-attribute root causes.
    \item Apriori (APR) is a popular frequent pattern mining algorithm~\cite{han2011data}. Ahmed et al.~\cite{ahmed2017detecting} and Lin et al.~\cite{lin2020fast} take association rules of abnormal leaf attribute combinations as root causes, and they use Apriori and \textit{confidence} to mine association rules.
    \item HotSpot+GRE~\cite{sun2018hotspot} (HS) uses Monte Carlo tree search (MCTS) to search the set of attribute combinations with the highest \textit{potential scores}. We adapt the original HotSpot for derived measures according to GRE.
    \item MID~\cite{gu2020efficient} searches for the attribute combinations that maximize their objective function, which is similar to that in iDice~\cite{lin2016idice}, and uses a heuristic based on entropy to speed up the search. 
    As we found that their objective function is limited to their scenario (\# issue reports) and performs poorly with general multi-dimensional data, we replace their objective function with our GPS.
    \item ImpAPTr~\cite{wang2020impaptr,rong2020locating} (IAP) search for attribute combinations that maximize \textit{impact factor} and \textit{diversity factor} with BFS (breath-first search). 
    Since ImpAPTr only ranks attribute combinations rather than decide which are the root-cause attribute combinations, we take the top-$n\_ele$ ranked attribute combinations as root-cause attribute combinations. 
    Besides, the original impact factor works for decreasing measure values, and thus, we modify it by deciding the sign of impact factor adaptively for each fault.
\end{itemize}
We do not compare with iDice~\cite{lin2016idice} due to its inferior performance in our scope according to our previous version~\cite{li2019generic} and MID~\cite{gu2020efficient}.
We set $\delta=0.9$ for all cases.
The parameters of other algorithms are set following the original papers.
Note that all approaches except ImpAPTr have no idea about $n\_ele$ or $cuboid\_layer$ of the faults.

\begingroup
\setlength{\tabcolsep}{4pt} %
\renewcommand{\arraystretch}{1} %
\begin{table}[htb]
\captionsetup{skip=0pt}
\caption{Time Usage Comparison of Forecast Methods}
\footnotesize
\begin{tabular}{llll}
\toprule
\textbf{Algorithm}      & MA                 & Period~\cite{lee2012threshold} & ARIMA~\cite{bhagwan2014adtributor} \\ \midrule
\textbf{Time Usage ($\mu s$)} & $6.21(\pm2.13)$ &    $28.7(\pm3.95)$                            &  $38672(\pm32312)$     \\ \bottomrule
\end{tabular}
\label[table]{tbl:forecast-time-usage}
\end{table}
\endgroup

In our evaluation, we always apply MA (moving average) for forecasting.
Specifically speaking, we calculate the forecast value of a leaf attribute combination $e$ at a specific time point $t_0$ by averaging the real values of $e$ at $t_{-10}, t_{-9}, ..., t_{-1}$.
We choose MA because MA is one of the simplest forecast algorithms and costs little time.
We present the time usage for a single leaf attribute combination of several algorithms used by existing work in \cref{tbl:forecast-time-usage}.

%% file: evaluation/overall_performance.tex
\subsection{RQ1: Effectiveness in Root Cause Localization}
\label{sec:overall-performance}

\begin{figure}[htb]
\captionsetup{skip=0pt}
\centering
\includegraphics[width=0.8\columnwidth]{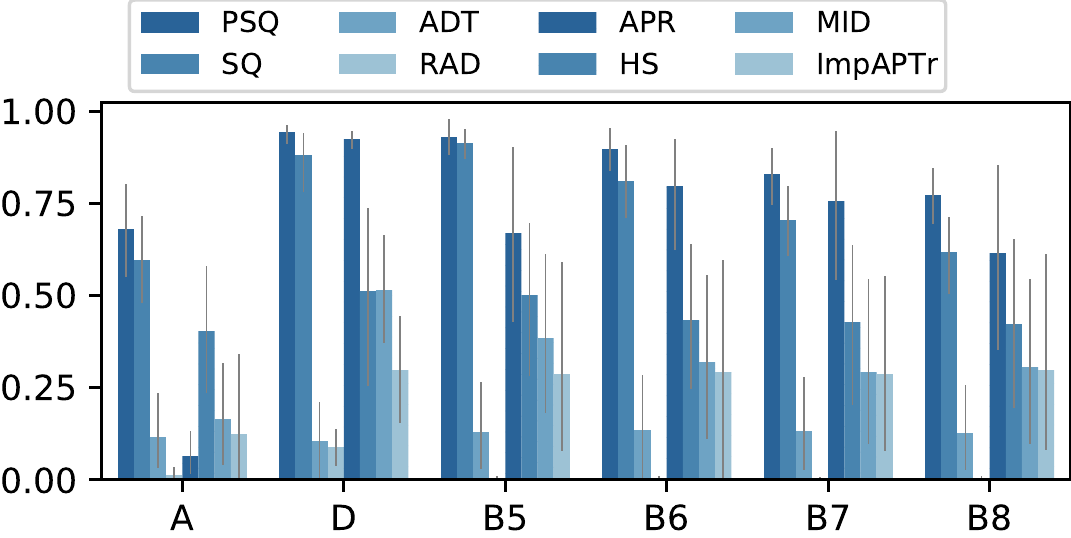}
    \caption{The F1-scores on simulated datasets.}
    \label{fig:performance-dataset}
\end{figure}

\begingroup
\setlength{\tabcolsep}{2pt} %
\renewcommand{\arraystretch}{1} %
\begin{table}[t]
\captionsetup{skip=0pt}
\centering
\caption{The $p$-value (by $t$-test) and effect size (by Cohen’s $d$~\cite{cohen1988spa}) of every baseline across all datasets}
\footnotesize
\begin{tabular}{llllllll}
\toprule
            & SQ      & ADT     & RAD     & APR     & HS      & MID     & IAP     \\ \midrule
p-value     & 7.9e-03 & 1.1e-41 & 6.2e-65 & 3.8e-04 & 1.9e-12 & 1.7e-18 & 3.3e-18 \\
effect size & 0.52    & 4.28    & 7.55    & 0.71    & 1.53    & 2.05    & 2.03  \\
\bottomrule
\end{tabular}
\label{tbl:p-value}
\end{table}
\endgroup

\input{evaluation/overall_performance_table}

As shown in \cref{fig:performance-dataset}, on average of different $n\_element$ and $cuboid\_layer$ settings, \ours{} achieves the highest performance in all datasets and outperforms the baselines significantly.
By further calculation, the F1-score of \ours{} outperforms the baselines (excluding Squeeze) by 32.89\% at least on average of all settings.
The improvement is significant according to \cref{tbl:p-value}.
Moreover, \ours{} is more robust and performs well consistently in different situations.
As shown in \cref{tbl:overall-performance-a-d} and \cref{tbl:overall-performance-b}, despite of different $n\_elements$, $cuboid\_layer$, and dataset settings \ours{} achieves good performance, and \ours{} outperforms all the other baselines in most (31 out of 54) settings.

\begin{figure}
\centering
\includegraphics[width=\columnwidth]{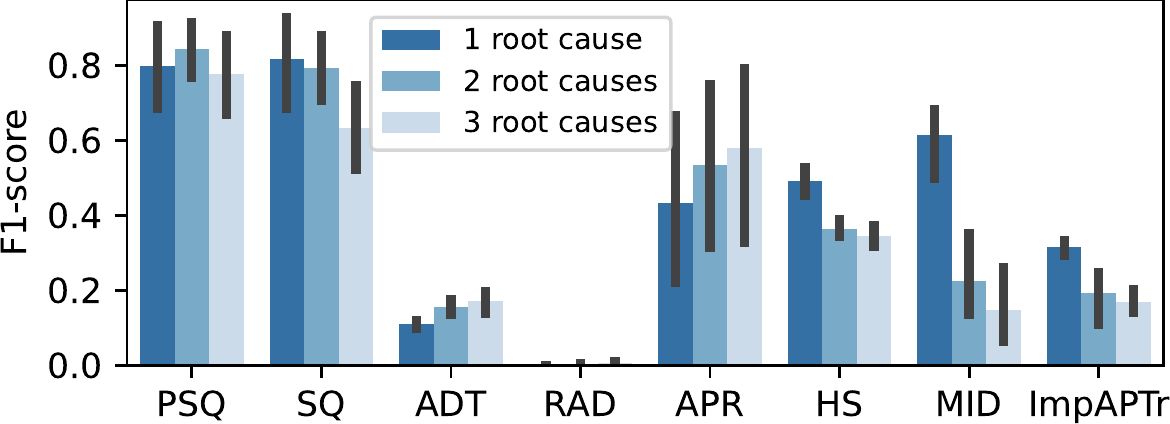}
\caption{The F1-scores with different numbers of root causes}
\label{fig:performance-multiple-root-causes} 
\end{figure}

In \cref{fig:performance-multiple-root-causes}, we present the F1-scores with different numbers of root causes on the simulated fault datasets.
The results show that \ours{} can achieve consistently good performance even if there is more than one root cause.

\begin{figure}[htb]
\captionsetup{skip=0pt}
\centering
\includegraphics[width=0.8\columnwidth]{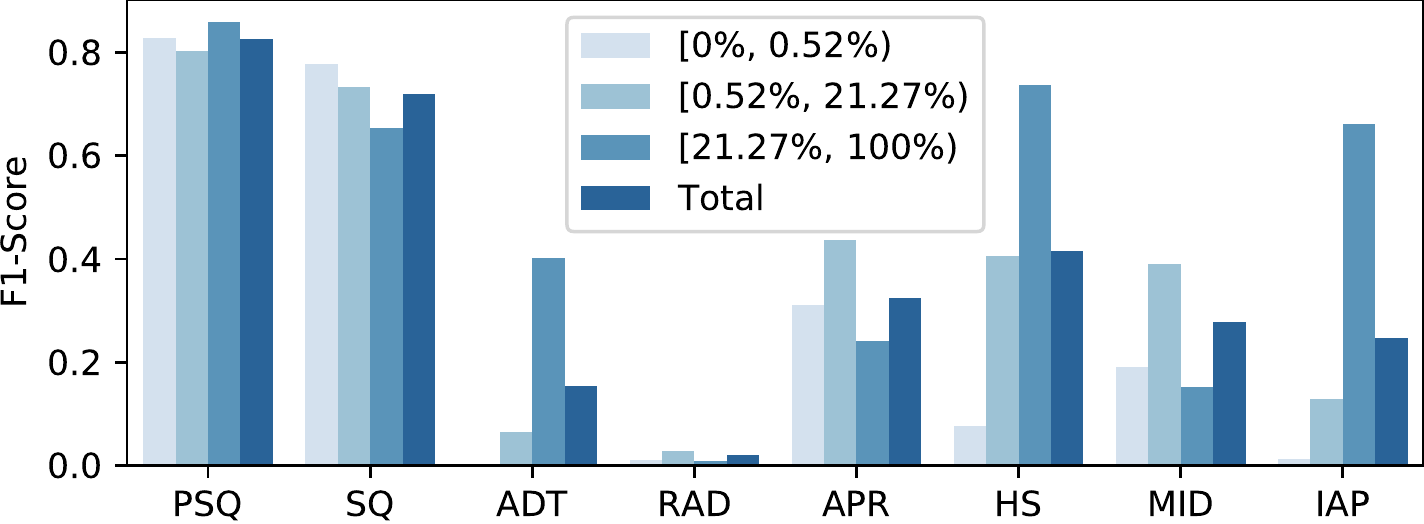}
    \caption{The F1-scores of faults with different anomaly magnitudes}
    \label{fig:performance-magnitude}
\end{figure}

Notably, \ours{} performs well consistently regarding different \textit{anomaly magnitudes}.
Formally speaking, the anomaly magnitude of a fault is $\frac{|\sum_{e\in LE(\emptyset)}(v(e)-f(e))|}{\sum_{e\in LE(\emptyset)}f(e)|}$, and it denotes the relative magnitude of the abnormal ﬂuctuation on the overall measure value.
As shown in \cref{fig:performance-magnitude}, \ours{} and Squeeze achieve high performance in spite of anomaly magnitudes.
However, the performance of Adtributor, HotSpot, and ImpAPTr varies largely with different anomaly magnitudes.
Though the performance of Apriori, R-Adtributor, and MID is relatively stable, their F1-scores are not high enough.

\ours{}, as well as many most other baselines, relies on forecast values.
As shown in \cref{tbl:overall-performance-b}, \ours{} performs worse as the forecast residual goes large (from $\mathcal{B}_1$ to $\mathcal{B}_4$, as shown in \cref{tbl:datasets}), but \ours{} consistently outperforms others even in the worst case.
The performance of Apriori is not monotonic to the forecasting residuals because its parameter setting is sensitive to the forecasting residual.
Some approaches like HotSpot seem affected less by forecasting residuals, but they perform poorly.
It worth noting that \ours{} is much more insensitive to forecasting residuals than our previous version, Squeeze, which further demonstrates the effectiveness of our extension (the probabilistic clustering method in \cref{sec:probabilistic-clustering}). 

In those settings where \ours{} does not achieve the best performance, either Squeeze (3 out of 54), Apriori (14 out of 54), HotSpot (3 out of 54), or ImpAPTr (2 out of 54) achieve the best performance.
However, the baseline approaches are not general and robust enough to perform well in different situations.
Adtributor only localizes root causes of the first-layer cuboids, and thus, it cannot work at all settings where $cuboid\_layer \neq 1$.
Its \textit{explanation power} is sensitive to the impact of attribute combinations (\ie{}, how much data is specified by attribute combinations), and thus it performs badly when anomaly magnitudes are small.
Note that faults with root causes in deeper cuboids often have smaller anomaly magnitudes.
Hence the performance of Adtributor decreases as $cuboid\_layer$ increases.
Although R-Adtributor localizes root causes in any cuboids, it is hard for R-Adtributor to decide when the recursion should terminate, and thus it works poorly in most settings.
Although Apriori achieves the best performance in several settings, it also works extremely badly in some settings (\eg{}, all settings on $\mathcal{A}$, $cuboid\_layer=3$ on $\mathcal{B}_1$) due to its sensitivity to parameters.
The \textit{hierarchical pruning strategy} could wrongly prune the right search path, and thus it performs poorly when $n\_ele$ or $cuboid\_layer$ is large.
Though both MID and ImpAPTr are designed for and limited to particular measures (the number of issue reports and success rate respectively), we have adapted them in our experiments.
However, they still suffer from other limitations.
The searching strategies of MID and ImpAPTr do no consider multiple root causes, and thus, they perform poorly when $n\_ele>1$.
Moreover, MID's heuristic strategy is limited to their scenario, and thus, when root-cause attribute combinations are in deeper cuboids, it is harder for MID to search it.
The \textit{contributor power} in ImpAPTr is sensitive to the impact of attribute combinations like \textit{explanation power}, and thus it is inappropriate for faults with small anomaly magnitudes.
While the baseline approaches suffer from these limitations, \ours{} is able to consistently achieve good performance in different situations.

\begin{conclusion}
    \ours{} is more general and robust to perform consistently well in different situations.
\end{conclusion}

\ours{} underperforms other baselines, especially Apriori, in some settings when the forecasting residuals are large (\ie{}, in datasets $\mathcal{B}_3$ and $\mathcal{B}_4$), or $n\_elements$ and $cuboid\_layers$ are large.
It is mainly because, in such cases, the bottom-up clustering step is affected by the noises.
For example, some normal leaf attribute combinations could be grouped into a cluster due to the large forecasting residuals.
To reduce the influence of noises, we apply probabilistic clustering in \ours{}.
As a result, as shown in \cref{tbl:overall-performance-b}, the larger the forecasting residuals are, the more \ours{} outperforms Squeeze.
In general, \ours{} outperforms Squeeze in most settings (50 out of 54) and the F1-score of \ours{} outperforms Squeeze by over 30\% in 11 out of 54 settings, and the improvement can be up to 113.7\%.
As the forecasting residuals increase from $\mathcal{B}_1$ to $\mathcal{B}_4$, the improvement of \ours{} over Squeeze increases from 1.96\% to 25.22\%, and the effect sizes (Cohen's $d$~\cite{cohen1988spa}) are $0.27$, $0.67$, $0.95$ and $1.10$ respectively.
Moreover, according to \cref{tbl:p-value}, the improvement is significant.
By further calculation, the F1-score of \ours{} outperforms Squeeze by 11.73\% on average.

\begin{conclusion}
    \ours{} significantly outperforms our previous version by reducing the influence of noises.
\end{conclusion}

\begin{figure}
    \centering
    \includegraphics[width=\columnwidth]{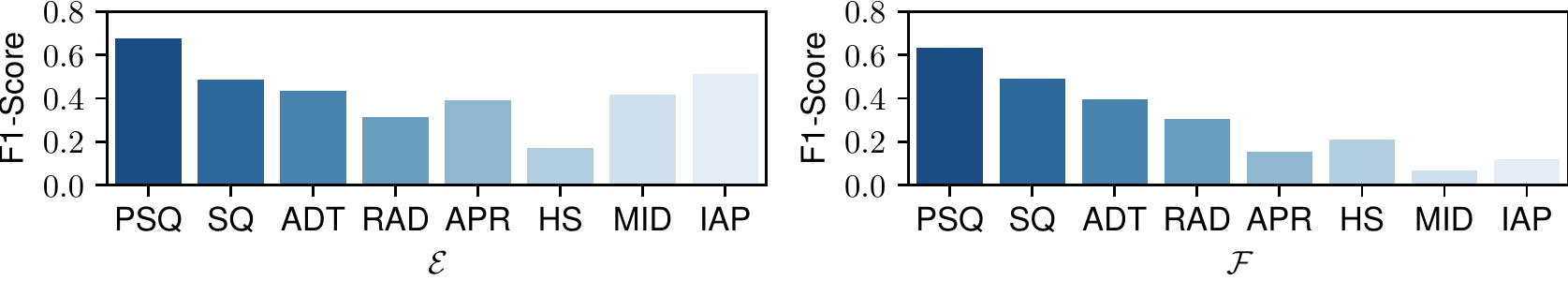}
    \caption{The F1-scores on datasets \ds{E} and \ds{F}}
    \label{fig:performance-trian-ticket}
\end{figure}

In \cref{fig:performance-trian-ticket}, we present the F1-scores on \ds{E} and \ds{F}.
The results show that \ours{} outperforms the baselines, including Squeeze, and achieves the best performance.
Compared with the simulated faults (i.e., \ds{A}, \ds{B} and \ds{D}), the performance of \ours{} slightly degrades on the inject faults.
It is probably because the data generation of \ds{E} and \ds{F} do not assume GRE holds at all.

\begin{conclusion}
    \ours{} can achieve good performance and outperforms the baselines in realistic faults.
\end{conclusion}

%% file: evaluation/overall_performance_table.tex
\begingroup
\setlength{\tabcolsep}{1.8pt} %
\renewcommand{\arraystretch}{1} %
\begin{table}[htb]
\captionsetup{skip=0pt}
\footnotesize
\caption{Overall performance comparison on \ds{A} and \ds{D}.}
\centering
\begin{tabular}{crrrrrrrrrrr}
    \toprule%
    \multicolumn{2}{c}{\textbf{F1{-}Score}}&\multicolumn{10}{c}{\textbf{n\_element,cuboid\_layer}}\\%
    \textbf{}&\textbf{Algo.}&\textbf{total}&\textbf{1,1}&\textbf{1,2}&\textbf{1,3}&\textbf{2,1}&\textbf{2,2}&\textbf{2,3}&\textbf{3,1}&\textbf{3,2}&\textbf{3,3}\\%
    \hline%

\multirow{12}{*}{${\mathcal{A}}$}
&\greycell{}PSQ&\greycell{}\textbf{0.68}&\greycell{}\textbf{0.97}&\greycell{}\textbf{0.93}&\greycell{}\textbf{0.70}&\greycell{}0.73&\greycell{}\textbf{0.67}&\greycell{}\textbf{0.67}&\greycell{}0.57&\greycell{}0.60&\greycell{}0.28\\

& SQ &0.60&0.94&0.75&0.52&0.68&0.65&0.57&0.37&\textbf{0.60}&\textbf{0.29}\\

&ADT&0.12&0.37&0.00&0.00&0.36&0.00&0.00&0.32&0.00&0.00\\

& RAD&0.01&0.06&0.00&0.00&0.00&0.01&0.00&0.00&0.00&0.05\\

&APR&0.06&0.31&0.03&0.00&0.05&0.02&0.07&0.06&0.03&0.00\\

&HS&0.40&0.71&0.35&0.24&\textbf{0.88}&0.27&0.01&\textbf{0.58}&0.32&0.25\\

&MID&0.16&0.69&0.32&0.03&0.00&0.23&0.01&0.00&0.19&0.01\\

&IAP&0.12&0.87&0.00&0.00&0.02&0.00&0.00&0.22&0.00&0.00\\

\hline 
\multirow{12}{*}{${\mathcal{D}}$}
&\greycell{}PSQ&\greycell{}\textbf{0.94}&\greycell{}0.95&\greycell{}\textbf{0.96}&\greycell{}\textbf{0.97}&\greycell{}\textbf{0.97}&\greycell{}\textbf{0.97}&\greycell{}\textbf{0.95}&\greycell{}0.84&\greycell{}\textbf{0.94}&\greycell{}\textbf{0.95}\\

&SQ&0.88&0.95&0.95&0.94&0.95&0.93&0.90&0.48&0.93&0.90\\

&ADT&0.10&0.16&0.00&0.00&0.32&0.00&0.00&0.46&0.00&0.00\\

& RAD&0.09&0.14&0.12&0.00&0.06&0.18&0.00&0.09&0.20&0.01\\

&APR&0.92&0.96&0.93&0.90&0.95&0.96&0.86&\textbf{0.96}&0.91&0.89\\

&HS&0.51&\textbf{0.99}&0.91&0.01&0.75&0.70&0.00&0.64&0.58&0.02\\

&MID&0.51&0.95&0.68&0.25&0.67&0.57&0.27&0.51&0.44&0.28\\

&IAP&0.30&0.41&0.43&0.04&0.50&0.34&0.04&0.64&0.26&0.01\\

\bottomrule%

    \label{tbl:overall-performance-a-d}
\end{tabular}    
\end{table}

\begin{table}[htb]
\captionsetup{skip=0pt}
\footnotesize
\caption{Overall performance comparison on ${\mathcal{B}_1}$, ${\mathcal{B}_2}$, ${\mathcal{B}_3}$ and ${\mathcal{B}_4}$.}
\centering
\begin{tabular}{crrrrrrrrrrr}

    \toprule%
    \multicolumn{2}{c}{\textbf{F1{-}Score}}&\multicolumn{10}{c}{\textbf{n\_element,cuboid\_layer}}\\%
    \textbf{}&\textbf{Algo.}&\textbf{total}&\textbf{1,1}&\textbf{1,2}&\textbf{1,3}&\textbf{2,1}&\textbf{2,2}&\textbf{2,3}&\textbf{3,1}&\textbf{3,2}&\textbf{3,3}\\%
    \hline%

\multirow{12}{*}{${\mathcal{B}_1}$}
&\greycell{}PSQ&\greycell{}\textbf{0.93}&\greycell{}\textbf{1.00}&\greycell{}\textbf{1.00}&\greycell{}\textbf{0.98}&\greycell{}\textbf{0.99}&\greycell{}\textbf{0.90}&\greycell{}0.89&\greycell{}0.96&\greycell{}0.80&\greycell{}0.85\\

&SQ&0.91&0.94&1.00&0.96&0.97&0.88&\textbf{0.91}&0.89&0.79&\textbf{0.87}\\

&ADT&0.13&0.29&0.00&0.00&0.43&0.00&0.00&0.45&0.00&0.00\\

& RAD&0.00&0.01&0.00&0.01&0.00&0.00&0.00&0.01&0.00&0.01\\

&APR&0.67&1.00&0.78&0.09&0.98&0.87&0.18&\textbf{0.96}&\textbf{0.93}&0.23\\

&HS&0.50&0.97&0.72&0.14&0.77&0.50&0.10&0.78&0.47&0.05\\

&MID&0.38&0.94&0.82&0.27&0.01&0.56&0.23&0.00&0.43&0.19\\

&IAP&0.29&1.00&0.00&0.00&0.86&0.00&0.00&0.71&0.00&0.00\\

\hline 
\multirow{12}{*}{${\mathcal{B}_2}$}
&\greycell{}PSQ&\greycell{}\textbf{0.90}&\greycell{}1.00&\greycell{}\textbf{0.99}&\greycell{}\textbf{0.93}&\greycell{}\textbf{1.00}&\greycell{}0.91&\greycell{}\textbf{0.80}&\greycell{}0.93&\greycell{}0.80&\greycell{}\textbf{0.74}\\

&SQ&0.81&0.91&0.97&0.93&0.99&0.78&0.75&0.78&0.53&0.66\\

&ADT&0.13&0.28&0.00&0.00&0.44&0.00&0.00&0.49&0.00&0.00\\

& RAD&0.00&0.00&0.00&0.00&0.00&0.00&0.01&0.01&0.00&0.01\\

&APR&0.79&0.97&0.91&0.36&1.00&\textbf{0.93}&0.49&\textbf{0.96}&\textbf{0.96}&0.58\\

&HS&0.43&0.97&0.60&0.18&0.73&0.43&0.03&0.68&0.27&0.00\\

&MID&0.32&0.96&0.86&0.26&0.01&0.42&0.11&0.01&0.18&0.05\\

&IAP&0.29&\textbf{1.00}&0.00&0.00&0.86&0.00&0.00&0.76&0.00&0.00\\

\hline 
\multirow{12}{*}{${\mathcal{B}_3}$}
&\greycell{}PSQ&\greycell{}\textbf{0.83}&\greycell{}0.88&\greycell{}\textbf{0.93}&\greycell{}\textbf{0.86}&\greycell{}0.95&\greycell{}0.85&\greycell{}\textbf{0.74}&\greycell{}0.93&\greycell{}0.74&\greycell{}\textbf{0.59}\\

&SQ&0.70&0.61&0.93&0.84&0.83&0.75&0.69&0.66&0.49&0.53\\

&ADT&0.13&0.28&0.00&0.00&0.41&0.00&0.00&0.50&0.00&0.00\\

& RAD&0.00&0.00&0.00&0.00&0.00&0.00&0.01&0.01&0.01&0.00\\

&APR&0.75&\textbf{1.00}&0.88&0.24&\textbf{0.99}&\textbf{0.93}&0.38&\textbf{0.97}&\textbf{0.96}&0.44\\

&HS&0.43&0.98&0.62&0.08&0.76&0.39&0.02&0.71&0.27&0.00\\

&MID&0.29&0.93&0.83&0.22&0.01&0.39&0.09&0.00&0.14&0.03\\

&IAP&0.29&0.99&0.00&0.00&0.84&0.00&0.00&0.74&0.00&0.00\\

\hline 
\multirow{12}{*}{${\mathcal{B}_4}$}
&\greycell{}PSQ&\greycell{}\textbf{0.77}&\greycell{}0.69&\greycell{}\textbf{0.89}&\greycell{}\textbf{0.76}&\greycell{}0.91&\greycell{}\textbf{0.80}&\greycell{}\textbf{0.72}&\greycell{}0.92&\greycell{}0.68&\greycell{}\textbf{0.58}\\

&SQ&0.62&0.32&0.86&0.75&0.69&0.70&0.65&0.58&0.46&0.54\\

&ADT&0.13&0.27&0.00&0.00&0.37&0.00&0.00&0.51&0.00&0.00\\

& RAD&0.00&0.00&0.00&0.00&0.00&0.00&0.00&0.02&0.00&0.01\\

&APR&0.61&\textbf{1.00}&0.72&0.05&\textbf{1.00}&0.78&0.09&\textbf{0.96}&\textbf{0.81}&0.12\\

&HS&0.42&1.00&0.71&0.10&0.71&0.30&0.02&0.70&0.26&0.00\\

&MID&0.31&0.97&0.84&0.25&0.00&0.41&0.09&0.00&0.15&0.05\\

&IAP&0.30&1.00&0.00&0.00&0.93&0.00&0.00&0.74&0.00&0.00\\
\bottomrule%

    \label{tbl:overall-performance-b}
\end{tabular}    
\end{table}
\endgroup

%% file: evaluation/performance_exrc.tex
\subsection{RQ2: Effectiveness in Determining External Root Causes}
\input{evaluation/erc_performance.tex}

Our method of determining external root causes works well on different datasets.
We randomly select and eliminate some attributes in the original datasets to emulate external root causes.
If a root cause contains any attribute that is eliminated, it becomes an external root cause.
As shown in \cref{tbl:external-root-cause}, \ours{} successfully determines external root causes with high F1-scores in almost all settings.
In 110 out of 135 settings, the $ExRC\_F1{-}Score$ of \ours{} achieves over 0.80.
By further calculation, the $ExRC\_F1{-}Score$ of \ours{} achieves 0.90 on average.
To the best of our knowledge, there is not any existing approach determining external root causes, and thus we do not compare with existing approaches. 
In some cases, the F1-scores are relatively low (\eg{}, 0.60).
The main reason is that there could be high-GPS attribute combinations even if the exact root causes are external due to the correlation among attributes.

\begin{conclusion}
    \ours{} can effectively determine external root causes.
\end{conclusion}

%% file: evaluation/erc_performance.tex
\begingroup
\setlength{\tabcolsep}{2.5pt} %
\renewcommand{\arraystretch}{1} %
\begin{table}[htb]
\captionsetup{skip=0pt}
\caption{Effectiveness on Determining External Root Cause}
\footnotesize
\centering
\begin{tabular}{crrrrrrrrrrr}
\toprule
\multicolumn{4}{c}{\textbf{ExRC\_F1{-}Score}}&\multicolumn{8}{c}{\textbf{n\_element,cuboid\_layer}}\\%
&$d_{ex}$&\textbf{total}&\textbf{1,1}&\textbf{1,2}&\textbf{1,3}&\textbf{2,1}&\textbf{2,2}&\textbf{2,3}&\textbf{3,1}&\textbf{3,2}&\textbf{3,3}\\%
\hline%
\multirow{3}{*}{$\mathcal{A}$}&1&0.78&0.65&0.73&0.95&0.91&0.82&0.92&0.61&0.84&0.62\\%
&2&0.82&0.67&0.97&0.95&0.78&0.84&0.93&0.72&0.89&0.60\\%
&3&0.92&0.97&0.96&0.96&0.88&0.88&0.96&0.75&0.95&0.94\\%
\hline
\multirow{3}{*}{$\mathcal{B}_1$}&1&0.91&1.00&0.97&0.96&0.92&0.72&0.97&0.86&0.82&0.93\\%
&2&0.99&1.00&1.00&0.99&1.00&0.99&1.00&0.99&0.94&0.98\\%
&3&1.00&1.00&1.00&0.99&1.00&1.00&1.00&1.00&1.00&1.00\\%
\hline
\multirow{3}{*}{$\mathcal{B}_2$}&1&0.89&1.00&0.99&0.94&0.87&0.85&0.89&0.80&0.73&0.90\\%
&2&0.99&1.00&0.99&1.00&1.00&0.99&0.98&0.97&0.98&1.00\\%
&3&0.98&1.00&0.99&1.00&0.97&0.99&0.97&0.95&0.98&1.00\\%
\hline
\multirow{3}{*}{$\mathcal{B}_3$}&1&0.85&0.81&0.97&0.89&0.87&0.84&0.95&0.78&0.68&0.86\\%
&2&0.93&0.81&1.00&0.94&0.88&0.96&0.95&0.88&1.00&0.97\\%
&3&0.95&0.91&1.00&0.95&0.91&0.99&0.95&0.91&0.99&0.98\\%
\hline
\multirow{3}{*}{$\mathcal{B}_4$}&1&0.77&0.74&0.91&0.93&0.76&0.67&0.77&0.68&0.70&0.75\\%
&2&0.89&0.71&0.94&0.93&0.79&0.96&0.93&0.78&0.98&0.95\\%
&3&0.89&0.71&0.96&0.93&0.75&1.00&0.96&0.81&0.97&0.95\\%
\bottomrule%
\multicolumn{12}{l}{The number of eliminated attributes is denoted as $d_{ex}$.}
\end{tabular}
\label{tbl:external-root-cause}
\end{table}
\endgroup

%% file: evaluation/efficiency.tex
\subsection{RQ3: Efficiency}
\label{sec:efficiency}

\begin{figure}[htb]
\captionsetup{skip=0pt}
    \centering
    \includegraphics[width=\columnwidth]{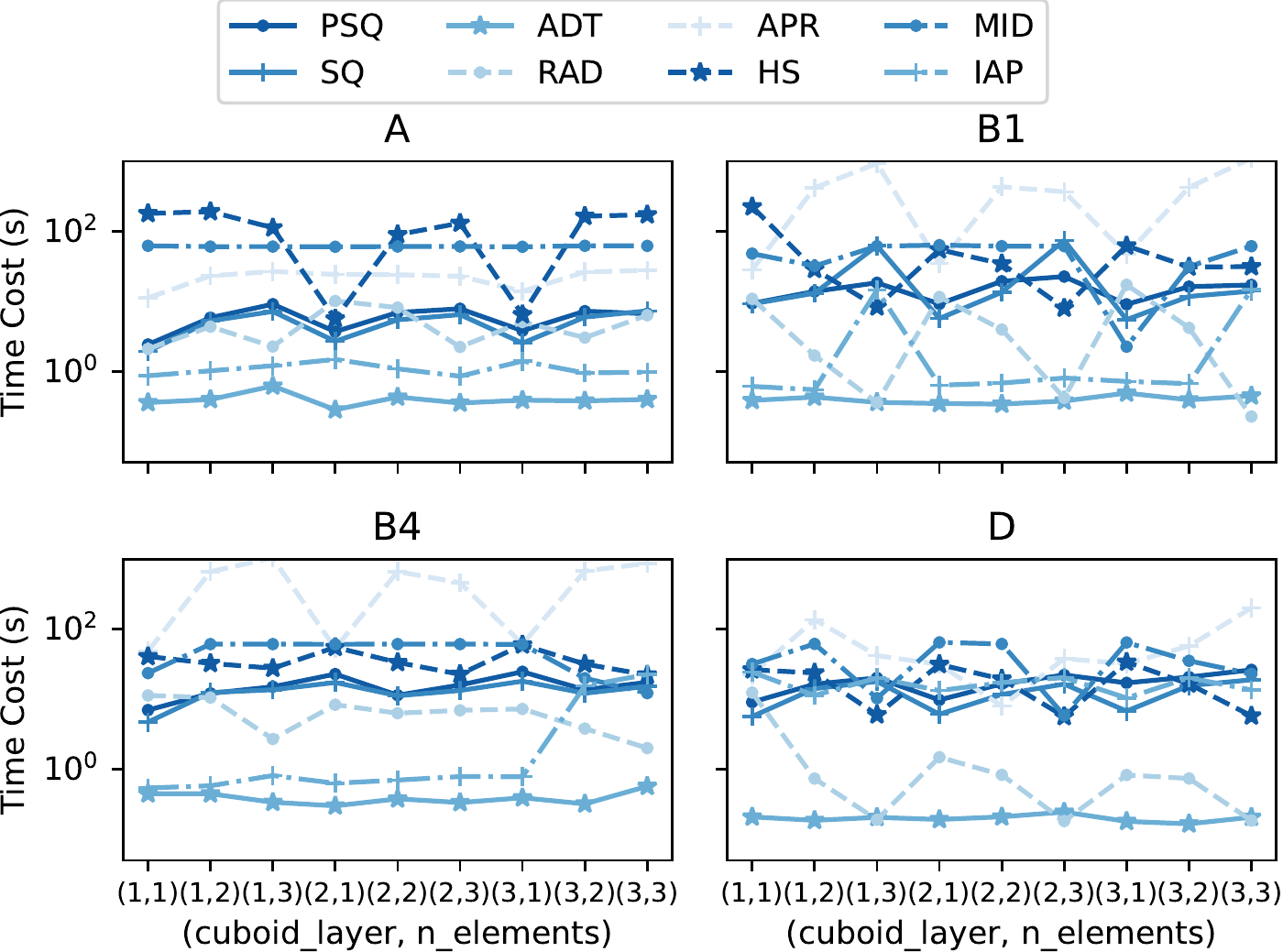}
    \caption{
    Running time comparison (for individual faults) of $\mathcal{A}, \mathcal{B}_1, \mathcal{B}_4, \mathcal{D}$.}
    \label[figure]{fig:running-time}
\end{figure}

We evaluate the efficiency of \ours{} by comparing its average time cost of each fault case with that of others.
We run every experiment on a server with 24 $\times$ Intel(R) Xeon(R) CPU E5-2620 v3 @ 2.40GHz (2 sockets) and 64G RAM.
All algorithms are implemented with Python utilizing matured libraries like Pandas and NumPy.
Experiments of all algorithms are conducted under the same condition.
In \cref{fig:running-time} we present the running times of $\mathcal{A}, \mathcal{B}_1, \mathcal{B}_4, \mathcal{D}$.
We do not present results of all datasets because all of $\{\mathcal{B}_i, i=1,2,3,4\}$ have similar results.
\ours{} costs only about ten seconds even in the worst cases.
It is efficient enough since measures are usually collected every minute or every five minutes.
HotSpot is sometimes as efficient as \ours{}, but sometimes it would cost more time.
Apriori costs hundreds of seconds, which is so slow that impractical.
Others can be fast, but they do not effectively localize root causes.

\begin{conclusion}
    \ours{} is efficient enough in practice to localize root causes for multi-dimensional data.
\end{conclusion}

%% file: evaluation/different_configuration.tex
\subsection{RQ4: Performance under Different Configurations}
\label{sec:different-configurations}

\begin{figure}[htb]
\captionsetup{skip=0pt}
\centering
    \includegraphics[width=0.6\columnwidth]{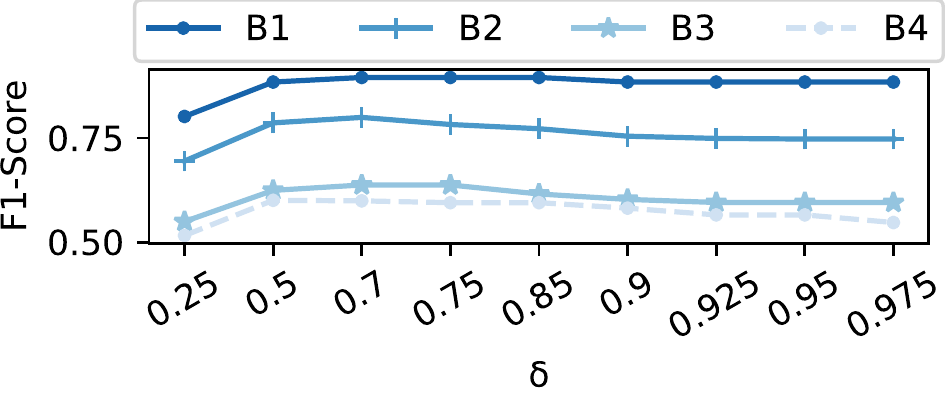}
  \caption{F1-Scores over different $\delta$ with $cuboid\_layer=3,n\_ele=3$.
  }
  \label{fig:ps-upper-bound}
\end{figure}

All the parameters in \cref{sec:methodology} are automatically configured except $\delta$, the GPS threshold.
We present the F1-scores of \ours{} under different $\delta$ in \cref{fig:ps-upper-bound} with  $cuboid\_layer=3, n\_ele=3$ in $\mathcal{B}_i,\>i=1,2,3,4$.
We only choose this setting because it is the hardest setting.
\ours{}'s performance does not change a lot as $\delta$ changes.
Results in \cref{tbl:overall-performance-a-d}, \cref{tbl:overall-performance-b} and \cref{sec:efficiency} also show that $\delta=0.9$ leads to good enough effectiveness and efficiency.
Since $\delta$ is the GPS threshold for early stopping and the distributions of GPS are not supposed to be related to datasets, it is reasonable to set $\delta$ near $0.9$ regardless of the specific dataset.

\begin{conclusion}
    \ours{} is robust to different configiurations.
\end{conclusion}

%% file: deployment.tex
\section{Success stories}
\label{sec:deployment}
We have successfully applied \ours{} in several large commercial banks and a top Internet company.
The results show that \ours{} can do great help for operators by rapidly and accurately localizing root causes.
In this section, we present some representative success stories.
For confidential reasons, some details are omitted or anonymized.
Compared with our previous version, we apply \ours{} in more production systems from different companies and collect more cases (\eg{}, case \RNum{4} and \RNum{5}).

\subsection{Case \RNum{1}: Insignificant Anomaly Magnitude}
\label{sec:case-1}
\begin{figure}[hbt]
\captionsetup{skip=0pt}
\centering
  \includegraphics[width=0.8\columnwidth]{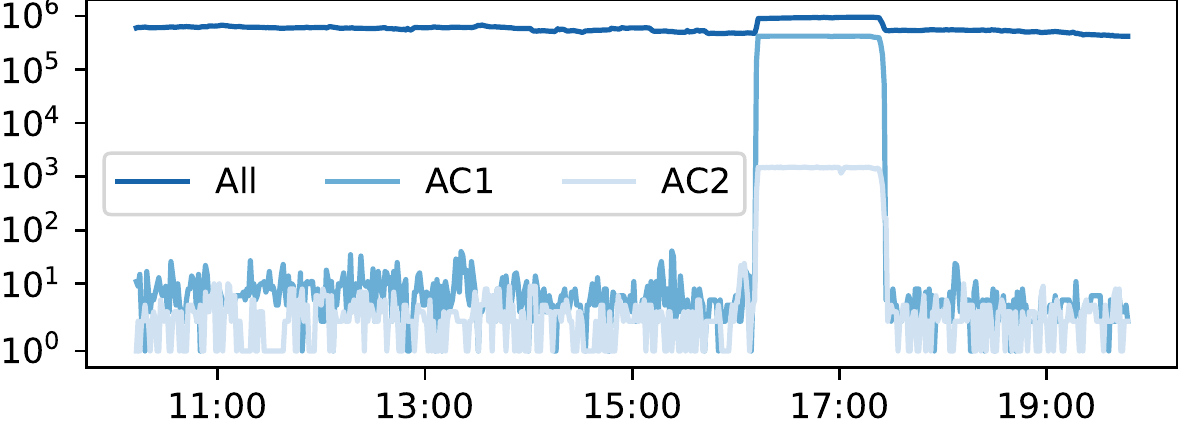}
 \caption{Measure values (in log-scale) along time of \textit{Case \RNum{1}}.
  $AC1$ is (user agent{=}uc{$\land$}idc{=}ih{$\land$}province{=}other), 
  and $AC2$ is (user agent{=}uc{$\land$}idc{=}is{$\land$}province{=}other).}
  \label[figure]{fig:case1}
\end{figure}

One day night, a fault occurred at a top Internet company.
The HTTP error counts suddenly burst, as shown in \cref{fig:case1}.
The attributes and the number of distinct values are listed as follows:
data center (11), province (7), ISP (6), user agent(22). 
A potential root cause is manually found by the operators, consisting of only one attribute combination ($AC1$ in \cref{fig:case1}), which took them one hour.

We retrospectively ran \ours{} over this system's logs, and in several seconds, we found more root causes: $AC1$ and $AC2$, as shown in \cref{fig:case1}. 
It is obvious that $AC2$ also has severe error bursts. 
$AC2$ is mistakenly ignored by the operators because it occupied only a small fraction of the total error counts.
Manual analysis apparently has difficulties in localizing root causes of anomalies with insignificant magnitudes. 
\ours{} would help the operators to notice such root causes efficiently. 

\subsection{Case \RNum{2}: Intra-System localization}
\label[section]{sec:real-case-intra-system}
One day from 9:00 to 11:00, the operators of a bank received many tickets and alerts and noticed that the API call success rate of a system suffered a severe drop. The search space is large, and the attributes and the number of distinct values are listed as follows: 
province (38), agency (815), server group (16), channel (4), server (339), code (4), status (2), service type (3). After two hours of fruitless manual root cause localization, the operators decided to just roll back the entire system to the last version, which happened to actually fix the issue. After the roll-back, it took another 2 hours for an inexperienced operator on duty to eventually find the root cause (there was a bug in the newly deployed version of the software for Service Type 0200020) based on the 2-hour logs during the fault.

Upon the request of the operators, we retrospectively ran \ours{} over this system's logs during the fault. 
\ours{} took a few seconds to report the root cause (Service Type=020020), which indicates exactly the software with a buggy version update. 
Had \ours{} been used immediately after the fault happened, operators could have localized the root cause much faster.

\begin{figure}[hbt]
\captionsetup{skip=0pt}
\centering
 \includegraphics[width=0.6\columnwidth]{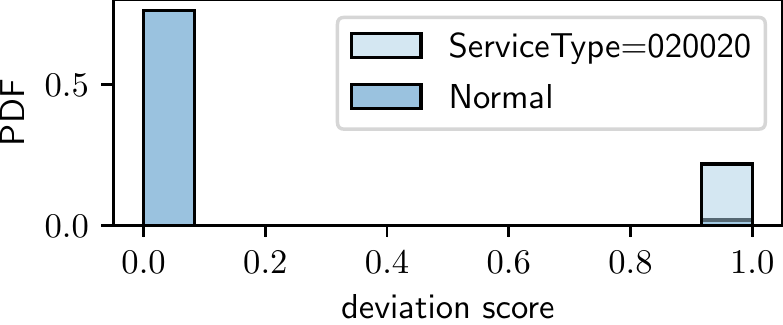}
 \caption{The histogram of deviation scores in \textit{Case \RNum{2}}.}
    \label[figure]{fig:case-1-clustering}
\end{figure}

\cref{fig:case-1-clustering} shows this case's results of deviation-score-based clustering. 
We can see that the deviation scores of all descended leaf attribute combinations of the root cause (Service Type=020020) are very close to each other. 
This to some extent confirms the generalized ripple effect.

\subsection{Case \RNum{3}: Inter-System localization}
\label{sec:real-case-inter-system}
\begin{figure}[hbt]
\captionsetup{skip=0pt}
\centering
    \includegraphics[width=0.8\columnwidth]{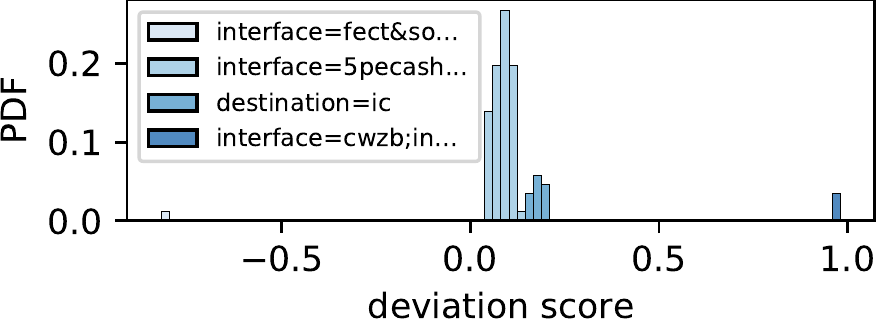}
 \caption{The root-cause attribute combinations and the histogram of deviation scores of their descent leaves in \textit{Case \RNum{3}}.}
    \label[figure]{fig:case2-deviation-score}
\end{figure}

One day, there was a burst of failures in a bank's transaction system.
There are many subsystems that communicate with each other by API (application programming interface) calls. 
The search space is also large, and the attributes and the number of distinct values are listed as follows:
source (13), source IP (66), destination (7), destination IP (10), interface (135). 
The operators located the root cause (destination=ic in \cref{fig:case2-deviation-score}) in ten minutes by manually analyzing many faulty traces (a series of API calls on different services to realizing on transaction is a trace~\cite{zhou2019latent,liu2020unsupervised,guo2020graphbased}).

Again, upon the request of the operators, we retrospectively ran \ours{} over this system's API call logs during the fault.
\ours{} localized the root causes, as shown in \cref{fig:case2-deviation-score}, in just several seconds.
It also confirms the generalized ripple effect because deviation scores of leaf attribute combinations that are descended from the same root-cause attribute combination are close to each other.
Note that \ours{} reports more root-cause attribute combinations than what the operators find.
The operators confirm that these additional root-cause attribute combinations are indeed valid: they are abnormal but are just affected by the \textit{ic} service due to the dependency among services.
Had \ours{} been actually used immediately after the issue, operators could have localized the root cause much faster (seconds vs minutes) and more accurately.
We also run some other algorithms on Case \RNum{3}~(see \cref{tbl:real-cases}).

\begingroup
\setlength{\tabcolsep}{3pt} %
\renewcommand{\arraystretch}{1} %
\begin{table}[hbt]
\captionsetup{skip=0pt}
 \caption{Qualitative comparison on industrial cases: whether the algorithm can find the true root cause. Some cases are missing due to deployment issues.}
 \centering
 \footnotesize
 \begin{tabular}{lcccccccc}
\toprule 
  true RC? & \ours{}& SQ & HS & APR  & ADT & RAD & MID & IAP\\
  \midrule
  Case \RNum{1} & Yes & Yes & No & No & No & No & No & No\\ 
  Case \RNum{3} & Yes & Yes & Yes & Yes & No & No & No & Yes\\ 
  Case \RNum{4} & Yes & No & No & Yes* & Yes & No & No & No\\ 
  Case \RNum{5} & Yes & No & No & Yes* & Yes & No & No & No\\
  \bottomrule
  \multicolumn{9}{l}{* Reporting too many false positives to be really helpful.}
\end{tabular}
\label{tbl:real-cases}
\end{table}
\endgroup

\subsection{Case \RNum{4} and \RNum{5}: External Root Causes}
\label{sec:real-case-external-root-causes}

One day at about 00:30, a fault occurred, and thus, a lot of transactions suffered long response latency at a system of a large commercial bank.
There are many attributes and attribute values:
transaction status (4), host IP (75), return code (275), transaction code (636), and MQ name (5).  
Therefore, manually localizing root cause attribute combinations is very challenging.

\begin{figure}[hbt]
\captionsetup{skip=0pt}
\centering
    \includegraphics[width=0.6\columnwidth]{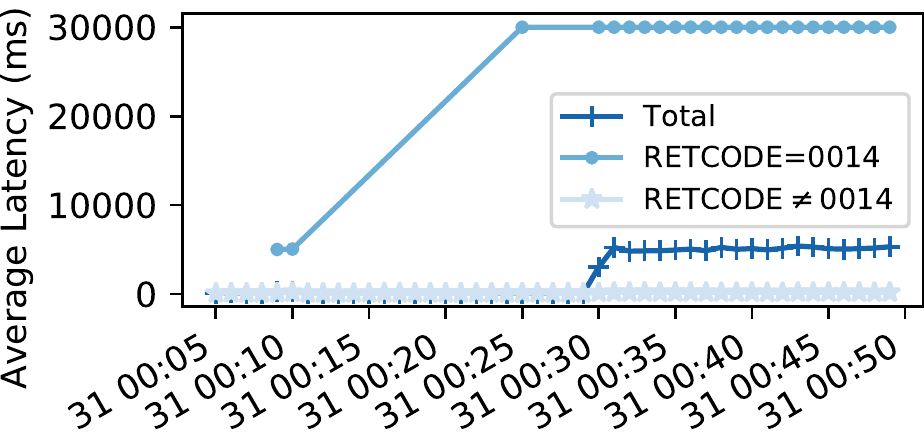}
    \caption{Case \RNum{4}. If the return code is 0014, then the average latency increases a lot; otherwise the average latency keeps steady.}
    \label{fig:case4}  
\end{figure}

We retrospectively ran \ours{} on this system's logs and successfully localize (return code = 0014) as the root cause (see \cref{fig:case4}).
Then the operator manually confirms this attribute combination directly indicates the exact underlying root cause.
We also compare \ours{} with other baselines in these cases and present the comparison in \cref{tbl:real-cases}.
Note that although Apriori successfully localizes the root cause, it also localizes 55 non-root-cause attribute combinations.

Initially, we did not take the two attributes, return code and transaction code, into consideration since there are too many attribute values of them.
Then \ours{} reported there might be an external root cause as $min\_GPS$ (refer to \cref{alg:external-root-cause}) was only $0.81$.
Moreover, all these approaches give invalid root causes:
\ours{}, HotSpot, Adtributor, MID and ImpAPTr localize almost all MQ names, R-Adtributor also localizes most MQ names combined with $transaction\>status=S$, and Apriori gives dozens of root causes.
Operators are not able to infer the underlying root cause from such results, and actually they can be misled.
Thus, we took all available attributes into consideration and then found the exact root cause.

\begin{figure}[hbt]
\captionsetup{skip=0pt}
\centering
\includegraphics[width=0.6\columnwidth]{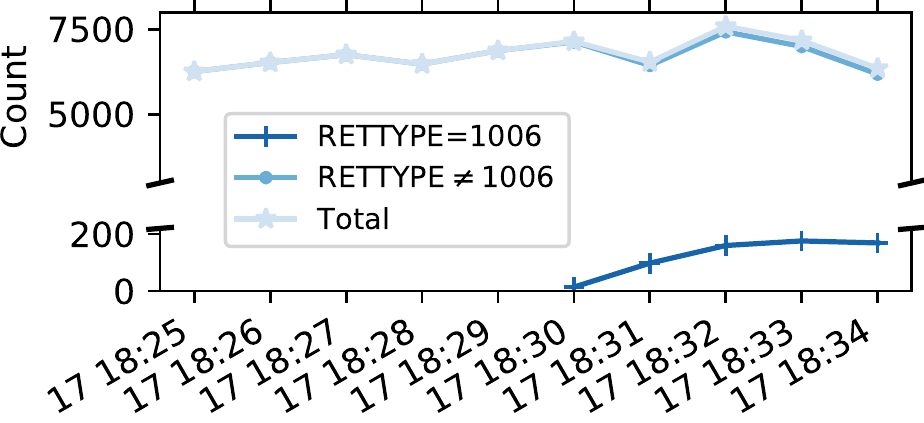}
    \caption{Case \RNum{5}. If the return type is 1006, then the transaction count increases a lot; otherwise the average latency keeps steady.}
    \label{fig:case5}    
\end{figure}

At another system of the same bank, another fault caused the transaction count to increase slightly and deviated from its normal state.
We also retrospectively ran \ours{} and other baselines, and the comparison is presented in \cref{tbl:real-cases}.
\ours{} accurately localizes the root cause attribute combination, ``RETTYPE=1006'' (see \cref{fig:case5}).
Note that Apriori localizes too many (17) non-root-cause attribute combinations again.
For both case \RNum{4} and \RNum{5}, Squeeze did not localize the exact root-cause attribute combinations due to the large forecasting residuals.
Similar to case \RNum{4}, we did not found the exact root cause until we were notified that there might be external root causes by \ours{} and considered more attributes.

%% file: discussion.tex
\section{Discussion}
\label{sec:discussion}

\subsection{Threats to Validity}
\label{sec:threats-to-validity}
The major threat to validity lies in the lack of real-world datasets.
However, given the difficulty in accessing real-world datasets, all the closely related works use simulated faults as well as our previous conference version, except iDice~\cite{lin2016idice} and MID~\cite{gu2020efficient}, which are both from Microsoft.
We think there are two reasons for the infeasibility of real-world datasets.
First, on the one hand, there are a limited number of faults in real-world production systems. Among them, the faults whose root causes can be indicated by root-cause attribute combinations is even less. On the other hand, in many cases, due to the lack of automated root-cause attribute combinations localization tools, the faults are diagnosed with the help of other monitoring data, such as logs and traces, and thus, the root-cause attribute combinations are missed in the failure tickets. Therefore, valid real-world cases that have ground-truth root-cause attribute combinations are somewhat rare. It could take years for engineers to collect hundreds of valid cases.
Second, multi-dimensional data are usually highly confidential. Unlike infrastructural metrics, such as CPU utilization and network throughput, the fields in the request logs that we use to build multi-dimensional data are usually sensitive, such as the dollar amount and the user’s location. The high risk also hinders companies from sharing real-world multi-dimensional data publicly.

To mitigate this threat to validity, first, we provide several real-world fault cases. 
Second, compared with the previous conference version, we added two new datasets (i.e., \ds{E} and \ds{F}), which were generated by realistic fault injection.

\subsection{Limitations}
\label{sec:limitations}

\textit{Root causes of multi-dimensional data} are not exact root causes of faults but only clues to them.
Nevertheless, localizing root causes for multi-dimensional data is important and helpful since it can direct further investigation right after faults occur.

\ours{}, as well as the previous approaches, relies on time-series forecasting.
To reduce the influence brought by inaccurate forecasting, especially when the real value is small, we introduce probabilistic clustering.
The experiment result shows that \ours{} is robust to forecasting residuals.

\ours{} focuses on only categorical attributes and cannot leverage numerical attributes directly, as well as most previous approaches~\cite{ahmed2017detecting,sun2018hotspot,rong2020locating,persson2018anomaly,lin2016idice,gu2020efficient,bhagwan2014adtributor,lin2020fast}.
We observe that numerical attributes are much less prevalent in practice (\eg{}, in the five companies studied in this paper). 
According to our interviews with some engineers, they choose not to record them because they are not sure how to use them for diagnosis.
We will work on supporting numerical attributes in the future.

\ours{} focuses on only numerical measures, as well as most previous approaches~\cite{bhagwan2014adtributor,persson2018anomaly,lin2016idice,sun2018hotspot,gu2020efficient,rong2020locating} except Apriori~\cite{ahmed2017detecting,lin2020fast}.
However, operators usually aggregate only numerical measures as time-series for monitoring and fault discovery in current industrial practice since categorical measures are not suitable for this. 
Thus operators are concerned about only numerical measures.

%% file: related_work.tex
\begingroup
\setlength{\tabcolsep}{2pt} %
\renewcommand{\arraystretch}{1} %
\begin{table*}[htb]
\captionsetup{skip=0pt}
\centering
\caption{Comparison of root cause localization approaches for multi-dimensional data \cref{sec:evaluation}}
\footnotesize
\begin{tabular}{cccccccc}    
\toprule
\multicolumn{1}{c}{\multirow{2}{*}{Approach}} & \multicolumn{2}{c}{Genericness}                                                                                                         & \multicolumn{3}{c}{Robustness}                                                                                                                                                                                               & \multirow{2}{*}{Efficiency} & \multirow{2}{*}{Search Strategy}\\ \cline{2-6}
\multicolumn{1}{c}{}                       & \begin{tabular}[c]{@{}l@{}}general\\ scenario\end{tabular} & \begin{tabular}[c]{@{}l@{}}measure\end{tabular} & \begin{tabular}[c]{@{}l@{}}any anomaly\\ magnitude\end{tabular} & \begin{tabular}[c]{@{}l@{}}determine external \\ root cause\end{tabular} & \begin{tabular}[c]{@{}l@{}}relying on parameter\\ fine-tuning\end{tabular} &                             \\
\midrule
Adtributor~\cite{bhagwan2014adtributor}                                        & no                                                                      & \greycell{}fundamental\&derived                                         & no                                                                & no                                                                        & \greycell{}no                                                                           & \greycell{}good             & top-down           \\
iDice~\cite{lin2016idice}                                        & no                                                                      & \#issue reports                             & no                                                               & no                                                                        & \greycell{}no                                                                           & depends  & top-down\\
Apriori~\cite{ahmed2017detecting,lin2020fast}                                        & \greycell{}yes                                                                     & \greycell{}fundamental\&derived                                         & \greycell{}yes                                                               & no                                                                        & yes                                                                          & depends             & bottom-up        \\
R-Adtributor~\cite{persson2018anomaly}                                       & \greycell{}yes                                                                     & \greycell{}fundamental\&derived                                         & \greycell{}yes                                                                & no                                                                        & yes                                                                          & \greycell{}good          & top-down              \\
HotSpot~\cite{sun2018hotspot}                                         & \greycell{}yes                                                                     & fundamental                                                  & no                                                                & no                                                                        & \greycell{}no                                                                           & depends         & top-down            \\                   
Squeeze~\cite{li2019generic}                                         & \greycell{}yes                                                                     & \greycell{}fundamental\&derived                                         & \greycell{}yes                                                               & no                                                                        & \greycell{}no                                                                           & \greycell{}good        & bottom-up\&top-down                \\
ImpAPTr~\cite{rong2020locating}                                        & no                                                                      & success rate                                          & no                                                                & no                                                                        & \greycell{}no                                                                           & \greycell{}good & top-down \\
MID~\cite{gu2020efficient}                                        & no                                                                      & \#issue reports                             & \greycell{}yes                                                               & no                                                                        & \greycell{}no                                                                           & depends               & top-down      \\
\greycell{}\ours{}                                        & \greycell{}yes                                                                     & \greycell{}fundamental\&derived                                         & \greycell{}yes                                                               & \greycell{}yes                                                                       & \greycell{}no                                                                           & \greycell{}good         &  bottom-up\&top-down              \\
\bottomrule                       
\end{tabular}
\label{tbl:related-works}
\end{table*}
\endgroup

\section{Related Work}
\label{sec:related-work}

Recently, many approaches focus on fault diagnosis in various contexts.
Most are different from ours~\cite{wu2018changelocator,lin2016log,zou2019empirical,chen2019empirical,chen2019continuous,zhou2019latent,liu2016focus,liu2020microhecl,guo2020graphbased,he2018identifying,li2021practical}.
On the one hand, we focus on root cause localization on multi-dimensional data.
On the other hand, some of these works use intuitive domain-knowledge-based empirical methods, while we propose a generic algorithm.

There are also several approaches focusing on localizing root causes for multi-dimensional data~\cite{bhagwan2014adtributor,persson2018anomaly,lin2016idice,sun2018hotspot,ahmed2017detecting,gu2020efficient,lin2020fast,rong2020locating}.
We compare them in \cref{tbl:related-works} in three aspects, \ie{}, genericness, robustness, and efficiency.
Some approaches are focusing on a specific scenario rather than generic multi-dimensional data.
For example, Adtributor~\cite{bhagwan2014adtributor} only cares about single-attribute root causes.
iDice~\cite{lin2016idice}, MID~\cite{gu2020efficient} and ImpAPTr~\cite{rong2020locating} utilize the special properties of specific types of measures (\ie{}, \#issue reports and success rate) and thus are not generic.
Many approaches~\cite{sun2018hotspot,lin2016idice,rong2020locating} filter attribute combinations by their impact (\eg{}, the percent of issue reports or transactions under them) and thus cannot handle insignificant anomalies.
Due to the design of termination conditions or pruning strategies, some approaches rely on parameter fine-tuning~\cite{persson2018anomaly,ahmed2017detecting,lin2020fast}, and the running time of some approaches~\cite{sun2018hotspot,lin2016idice,gu2020efficient,ahmed2017detecting,lin2020fast} varies in different faults.

Time series forecasting has been extensively studied, and there are many approaches.
Statistical approaches~\cite{chen2013providerside,zhang2018funnel,lee2012threshold,ma2018robust} make some statistical assumptions on the time series.
Supervised ensemble approaches~\cite{liu2015opprentice} try to ensemble statistical approaches in a supervised manner.
Recently, unsupervised deep-learning-based approaches~\cite{xu2018unsupervised,li2018robust} are making great progress.
The selection of appropriate forecasting algorithms is usually based on the nature of data~\cite{liu2015opprentice}.

There are many other studies on the analysis of multi-dimensional data.
A series of studies~\cite{tang2017extracting,lin2018bigin4,ding2019quickinsights} focus on identifying interesting patterns (a.k.a. insights) in multi-dimensional data.
Lumos~\cite{pool2020lumos} diagnoses metric regressions by ranking attributes by their importance after regression.
Castelluccio et al.~\cite{castelluccio2017automatically} focus on mining contrasting sets by statistical tests, which are attribute combinations with distinctive supports in different groups, to find attribute combinations related to a specific group of crashes.
Liu et al.~\cite{liu2016focus} debug high response time by mining distinctive conditions for high response time with the help of the decision tree algorithm.

%% file: conclusion.tex
\section{Conclusion}
\label{sec:conclusion}

Given the importance of root cause localization for multi-dimensional data, many approaches are proposed recently.
However, they are not generic or robust enough due to some limitations.
In this paper, we propose a more generic and robust approach, \ours{}.
\ours{} employs a novel ``bottom-up\&top-down'' searching strategy based on our proposed generalized ripple effect to achieves high efficiency without much loss of genericness and robustness.
Notably, this paper further extends our previous studies by a probabilistic clustering method and a method for determining external root causes.
We conduct extensive experiments on both simulated and injected faults.
The results show that the F1-score of \ours{} outperforms previous approaches by 32.89\% on average and consistently costs only about 10s.
Besides, the F1-score in determining external root causes reaches 0.90 on average.
Furthermore, case studies in several large commercial banks and an Internet company show that \ours{} can localize root causes much more rapidly and accurately than traditional manual analysis in practice.